# Title: Chirality Inversion and Giant Optical Activity in Scalable Self-Assembled Metamaterials

**Authors:** Antonio Carone[1,†,*], Aso Rahimzadegan[1,†], Birgül Kalkan[1], Thomas G. Parton[1], Tony Jin[1], Zihao Lu[2], Clemens Schmitt[1], Emeline Raguin[1], Bruno Frka-Petesic[2,3], Kevin Vynck[4], Silvia Vignolini[1,2,*].

**Affiliations:**
[1] Max Planck Institute of Colloids and Interfaces (MPIKG), Potsdam, Germany.
[2] Department of Chemistry, University of Cambridge, Cambridge, UK.
[3] International Institute for Sustainability with Knotted Chiral Meta Matter (WPI-SKCM$^2$), 1-3-1 Kagamiyama, Hiroshima University, Higashi-Hiroshima, Hiroshima, 739-8531, Japan.
[4] Institute of Light and Matter, Claude Bernard University Lyon 1, CNRS, Villeurbanne, France.

† These authors contributed equally to this work.
*Corresponding authors. Email: sv@mpikg.mpg.de; antonio.carone@mpikg.mpg.de

**Abstract**

Chiral materials manipulate light in ways that hold promise for sensing, imaging, and photonics, but producing such materials with strong and reproducible optical activity at large scale remains challenging. Here, we develop a scalable strategy to assemble hierarchical helicoidal metamaterials from metallic nanoparticles and chitin nanocrystals, a biosourced polysaccharide. Spontaneous co-assembly produces ordered chiral structures with strong plasmonic optical activity, reaching a dissymmetry factor of 0.55. The unprecedented ability to vary the helicoidal pitch allows control of the optical response, enabling inversion of optical handedness without changing the handedness of the template, effectively decoupling geometrical chirality from optical chirality. Remarkably, the gold nanoparticles can also be recovered from dry films and redispersed as colloids, introducing an appealing circularity concept for sustainable chiral plasmonic materials.

Chirality, defined as the property of an object being non-superimposable on its mirror image, is a fundamental geometrical property with profound implications across physics, chemistry, and biology. Its widespread importance has motivated the development of chiral nanomaterials (*1*), leading to several discoveries in biomedicine (*2*), enantiomer recognition (*3*, *4*), asymmetric catalysis (*5*), energy conversion (*6*), spin-selective transport (*7*) and optoelectronics (*8*). In this context, metallic chiral nanostructures (*9*) are particularly attractive because their localized resonances enable strong light-matter interactions and enhanced optical activity at subwavelength scales (*10–13*). In these nanostructures, the optical activity strongly depends on their geometry at the nanometer scale, which raises an important challenge for applications: tuning predictively their chiral architectures while keeping them manufacturable at scale. Lithography provides precise control but is costly and difficult to scale (*9*, *14*). Such approaches illustrate the trade-off between precise control of local geometry and the fabrication of extended chiral structures. Colloidal synthesis of chiral plasmonic nanoparticles has yielded systems with strong chiroptical activity, with g-factors in the range of 0.2-0.4 (*2*, *15–17*). However, these systems rely on stringent synthetic processes and high-energy morphologies that are susceptible to reshaping (*18*). An alternative is to use simpler, achiral nanoparticles, whose positions and orientations can be controlled to generate collective chirality. Self-assembly offers greater flexibility in designing chiral architectures, adding a good control over the optical activity, from individual nanoparticles to their collective organization. This has been demonstrated with DNA-origami assemblies, which can achieve nanometer-scale precision but remain limited in scalability and generally exhibit low g-factors, typically in the $10^{-2}$ range (*19–23*). However, in all these methods, spectral tunability, high reproducibility, scalability, and large dissymmetry factors (g ≥ 0.1) are rarely achieved simultaneously. Here, we co-assemble achiral gold bipyramids with chitin nanocrystals (ChNCs) to form large-area helicoidal metamaterials with pitches spanning one order of magnitude, from a few hundred nanometers to several micrometers. Such a range of tunability allows direct control of the relative orientation of neighboring particles, enabling fine control of the chiral plasmonic response. Three-dimensional tomography and an experimentally informed model relate nanoparticle density and orientation to the assembly's geometric chirality, while electromagnetic simulations explain its chiroptical response. We achieve a directly measured g-factor of 0.55 and show that changing the template pitch and plasmonic particle density can invert the optical handedness without reversing the left-handed ChNC template. These films are manufacturable at scale and can be disassembled on demand, enabling recovery of gold nanoparticles as stable colloids and introducing an element of material circularity. Together, these findings establish a scalable route to strong optical activity and reveal how chirality emerges from structure across length scales.

## Film fabrication and characterization

To produce the chiral metamaterial, we developed a co-assembly strategy combining gold bipyramids (AuBPs) and chitin nanocrystals (ChNCs) into helicoidal structures (Fig. 1A, figs. S1-S3, see Materials and Methods). In contrast to previous work with cellulose nanocrystals and plasmonic nanoparticles (*24–27*), ChNCs offer an advantage by forming an optically inactive chiral template that does not interfere with the plasmonic response (*28*, *29*). Moreover, compared with nanorods, AuBPs offer sharper plasmonic resonances (*30*), which is favorable for a strong chiroptical response. Under appropriate physicochemical conditions (pH, ionic strength, and concentration), ChNC suspensions spontaneously self-assemble into a chiral nematic phase (*31*). To enable the incorporation of well-dispersed AuBPs within the helicoidal template, we ensured that both colloidal particles possess positive surface charge (see Materials and Methods).

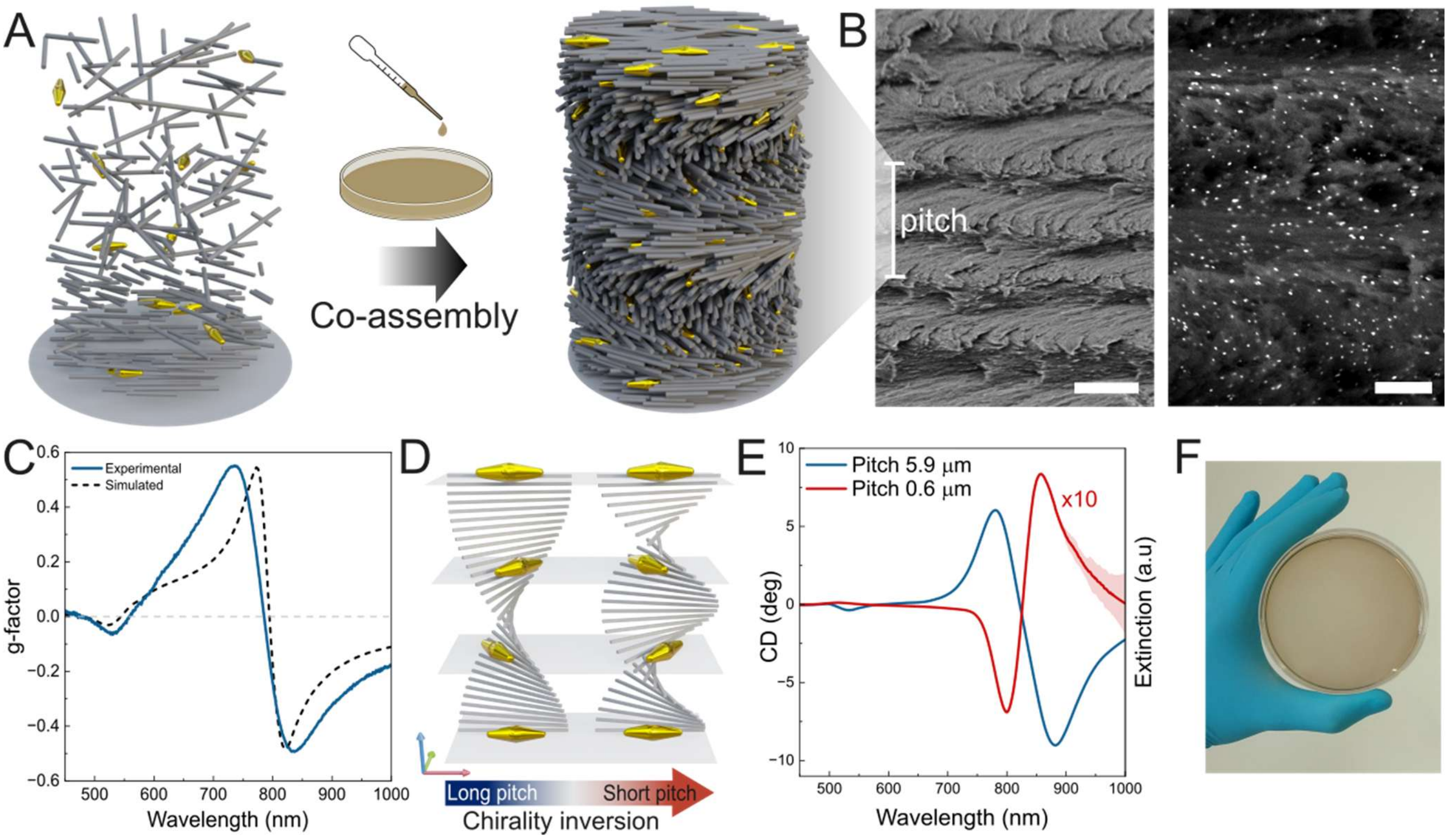

**Figure 1. Preparation, structure, and chiroptical properties of the plasmonic films.** (**A**) Schematic illustration of film fabrication composed of chitin nanocrystals (ChNCs) and gold bipyramids (AuBPs).

ChNCs are represented as gray rods and AuBPs as gold elongated particles. (**B**) Cross-sectional SEM images showing the periodic helicoidal structure of the film (left) and AuBP distribution, visualized by back-scattered electron imaging (right). Scale bars: 5 µm (left), 1 µm (right). (**C**) Experimental (solid line) and simulated (dashed line) g-factor from transmission measurements, reaching a maximum of 0.55. The bands are centered at the characteristic longitudinal (~780 nm) and transverse (~530 nm) resonances of the elongated nanoparticles (aspect ratio ≈ 3.2). Simulations were performed using a T-matrix approach on a subset of the *Tomographic Dataset* comprising ~47,000 identical AuBPs in a homogeneous chitin background. (**D**) Schematic illustrating how AuBPs can adopt mirror-related local arrangements in helicoidal templates with different pitches but the same left-handed chirality. (**E**) Circular dichroism spectra of films with different pitches containing AuBPs (aspect ratio ~3.2). The spectrum corresponding to a pitch of 0.6 µm was multiplied by a factor of 10 for better comparison. (**F**) Photograph of a film inside a 90 mm Petri dish.

To monitor co-assembly in situ, we confined the suspensions in capillaries or cuvettes and observed them by polarized optical microscopy. This revealed the periodic "fingerprint" texture of chiral nematic phases, along with the light absorption associated with the plasmonic particles (fig. S4). The suspensions with chiral nematic order were subsequently dried to produce solid structures retaining the helicoidal order. Scanning electron microscopy (SEM) of film cross-sections revealed well-formed helicoidal structures with the AuBPs dispersed inside the ChNC template (Fig. 1B and figs. S5, S6). We investigated a range of AuBP loadings between 5 and 150 particles/$\mu m^3$ and identified the latter as a soft upper limit, around which local phase separation into AuBP-rich droplets occurred (fig. S7). Transmission circular dichroism (CD) spectroscopy of the dry films revealed strong plasmonic chiroptical activity, with signal intensity depending on AuBP concentration (fig. S5) and the template pitch (here defined as a full 360° turn of the ChNCs). The highest measured dissymmetry factor (g-factor) was 0.55 (see Materials and Methods), observed for films with a nominal particle density of 150 particles/$\mu m^3$ and a pitch of 4-5 µm (Fig. 1C). To ensure the reliability of the measured values, we directly measured transmission under left- and right-circularly polarized illumination and averaged measurements at different sample orientations to minimize artifacts arising from linear dichroism and birefringence (see Materials and Methods) (*32–35*). No CD signal was detected in control films, such as pure helicoidal ChNC films or those containing AuBPs in a non-helicoidal template (fig. S8). This confirms that both the plasmonic component and the helicoidal organization are required to generate the chiroptical response. Transition matrix (T-matrix) simulations (*36*) further corroborate the plasmonic origin of the response and reproduce the measured chiroptical spectra (Fig. 1C and see Materials and Methods). The calculations use the experimentally reconstructed AuBP distribution as detailed in the following sections. By introducing AuBPs into a helicoidal template with a fixed handedness (left-handed in the case of ChNC), both right- and left-handed geometries of plasmonic nanoparticles can be achieved by changing the length of the helical pitch. A schematic illustrating this effect is shown in Fig. 1D, where elongated particles can adopt mirror-like chiral configurations of the same or opposite handedness to the cholesteric template. Such geometric considerations apply to any continuous helicoidal template that hosts discrete anisotropic inclusions. We observe an inversion of the CD signal when the pitch is reduced from 5.9 ± 0.6 µm to 0.6 ± 0.1 µm (Fig. 1E), although the chitin template remains left-handed. We found that both the intensity and spectral features of the CD response depend on AuBP loading and on the pitch of the ChNC template. In the following section, we discuss in more detail the distribution of the particles and its impact on the optical response.

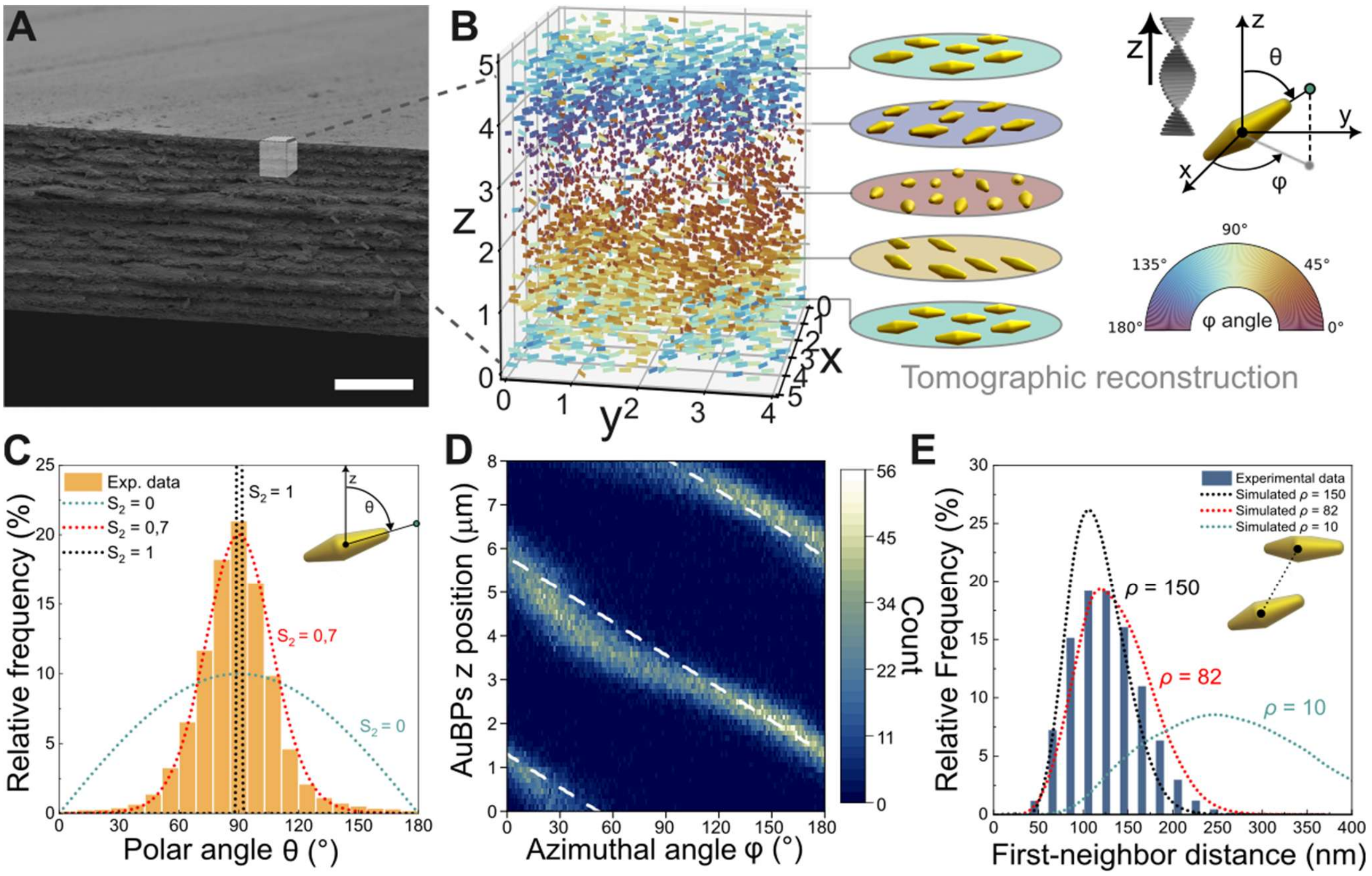

**Figure 2. Tomographic reconstruction.** (**A**) Schematic representation of the section of the film analyzed by FIB-SEM (overlay not to scale). Scale bar 10 µm. (**B**) Tomographic reconstruction from FIB-SEM of the AuBPs color-coded by their azimuthal angle ($\varphi$). The AuBPs are represented as rods with coordinates ($x$, $y$, $z$) and orientation ($\theta$, $\varphi$). Axes are in µm. (C) Polar angle ($\theta$) distribution for the AuBPs in the tomographic dataset, including angle distributions simulated using different $S_2$ order parameters. (**D**) Correlation diagram of the z position of AuBPs and their angle $\varphi$ obtained from the tomographic reconstruction. The dotted line represents the expected trend for a left-handed distribution defined as $\varphi_{(z)} = [\varphi_0 - 360° \times z/\text{pitch}] \bmod 180°$, with pitch = 9 µm and offset $\varphi_0 = 415°$. (**E**) First-neighbor distance distribution of the tomographic data. The dotted lines show the distribution obtained using a modeled tailored disorder point pattern with varying AuBP particle densities (N/µm$^3$).

## Tomographic reconstruction

We performed SEM cross-section analysis, which showed that AuBPs tend to align with the chitin template (fig. S6). However, these two-dimensional cross-sectional images remain limited in characterizing the three-dimensional (3D) structure of the material. To overcome this problem, we performed focused ion beam scanning electron microscopy (FIB-SEM) to reveal the 3D distribution and orientation of the AuBPs (Fig. 2A, see Materials and Methods). The tomographic analysis was conducted over a volume of ≈ 1600 µm$^3$ and it comprised more than $1.3 \times 10^5$ particles. The contrast between gold and chitin allowed us to determine the position coordinates and the orientation of the AuBPs in three dimensions. We represented each AuBP orientation and position in hemispherical coordinates, and hereafter refer to the reconstructed distribution as the "*Tomographic Dataset*". Figure 2B presents an inset of the tomographic reconstruction of AuBPs color-coded by the individual azimuthal angle $\varphi$. We found the AuBPs to predominantly lie perpendicular to the $z$-axis (Fig. 2C) and rotate around it (fig. S9), forming a left-handed helix guided by the local ChNC arrangement. Because of the centrosymmetric geometry of the AuBPs, $\varphi$ and $\theta$ were both defined within the [0, 180°] interval. The orientational order along the local director was quantified using the nematic order parameter $S_2$, which ranges from 0 for an isotropic distribution to 1 for a perfect parallel alignment (*37*). We calculated $S_2$ = 0.7 for the *Tomographic Dataset* (see Materials and Methods), consistent with the analysis of the $\theta$ angles (Fig. 2C). More informative is the correlation between the AuBP azimuthal angle φ and z position (Fig. 2D). The "descending" diagonal stripes in the correlation diagram confirm a left-handed rotation around the *z*-axis, with a half-turn period (P/2) of approximately 4.5 μm, matching both handedness and periodicity of the chitin matrix used to template the AuBPs (fig. S10). A comparison with a simulated distribution of AuBPs with the same pitch and $S_2$ parameter can be found in figure S11.

## Pitch-dependent inversion of chirality

The analysis of the *Tomographic Dataset* proved useful for rationalizing the geometry of the system. To extend this type of analysis to samples with different nanoparticle densities and helicoidal pitches, we developed an empirically informed model. We focused on the observed key features of the *Tomographic Dataset*: The AuBP positions and the orientations. We used two tomographic datasets at different concentrations of AuBPs to tailor a disordered point process that reproduces the positions of the particles, nearest-neighbor distance distribution (Fig. 2E), and pair correlation function (figs. S12 and S13). We refer to this empirically informed model as the *Helicoidal Model* (Fig. 3A). We then used the model to systematically study how pitch and particle density affect the chirality of the assembly. Previous studies of chiral plasmonic assemblies (*38*, *39*), together with our AuBP dimer simulations (fig. S14), indicated that the dihedral angle between elongated particles governs their chiroptical response. We therefore took the relative orientation between nearest-neighbor AuBPs separated along the z-axis to define a geometrical chirality descriptor $\chi_{\text{geom}} = (\hat{r}_{12} \cdot \boldsymbol{w}_1 \times \boldsymbol{w}_2)(\boldsymbol{w}_1 \cdot \boldsymbol{w}_2)$, where $\boldsymbol{w}_1$ and $\boldsymbol{w}_2$ are the orientation vectors of two neighboring particles, and $\hat{r}_{12}$ is the unit vector connecting them (fig. S15, see Materials and Methods) (*40*, *41*). We then calculated the mean geometrical chirality ($\bar{\chi}_{\text{geom}}$) across ≈1.3 × $10^5$ particles generated by the *Helicoidal Model* and mapped $\bar{\chi}_{\text{geom}}$ as a function of pitch and particle density (Fig. 3B). Figure 3B shows that the geometrical chirality inverts as a function of matrix pitch and particle density. This behavior arises because the pitch ($P$) and interparticle distance ($d$) determine the dihedral angle between neighboring particles, $\Delta\varphi = 2\pi d/P$. Changes in pitch or density therefore alter the dihedral angle and consequently the geometrical chirality. The dotted line marks the predicted inversion boundary, calculated by approximating $d$ using the mean nearest-neighbor distance (Eq. 13, see Materials and Methods). To systematically test this geometrical prediction experimentally, we developed a strategy to control the pitch length of the plasmonic films over a wide range. Although changing the ionic strength and pH can tune the pitch during assembly (*31*), the accessible range is limited. We previously reported that ChNCs from different biological sources exhibit remarkably different pitches under similar assembly conditions, ~4 μm for shrimp-derived ChNCs and ~400 nm for fungal-derived ChNCs (*42*) (fig. S1). We therefore developed a strategy that mixes shrimp- and fungal-derived ChNCs in controlled ratios, allowing continuous tuning of the pitch between those of the pure components. Films prepared at intermediate compositions exhibited pitches being the harmonic average weighted by the ChNC mass fractions (fig. S16). Using this approach, we prepared samples with varying pitch and particle density and measured their CD spectra. Figure 3B shows how the overlaid experimental optical handedness matches the predicted sign change in $\bar{\chi}_{\text{geom}}$. Representative spectra of samples P1-P3 are shown in Figure 3. Overall, the geometrical index predicts the handedness of the CD signals for most experimental samples. However, this analysis cannot predict the magnitude or spectral line shape of the response. For example, sample P2 (gray points and spectrum in Fig. 3B and 3C) or other samples near the inversion line exhibit a strong negative Cotton effect that cannot be explained by the geometrical analysis alone. To investigate this feature, we performed electromagnetic simulations on AuBP ensembles generated with the *Helicoidal Model*. We modeled the AuBPs as identical bicones with spherical tips in a homogeneous chitin background (fig. S17, see Materials and Methods). The T-matrix simulations reproduce the CD spectra of samples P1-P3 (Fig. 3C), indicating that the electromagnetic simulations capture the optical response of the structure and are suitable for exploring a broader pitch-density parameter space.

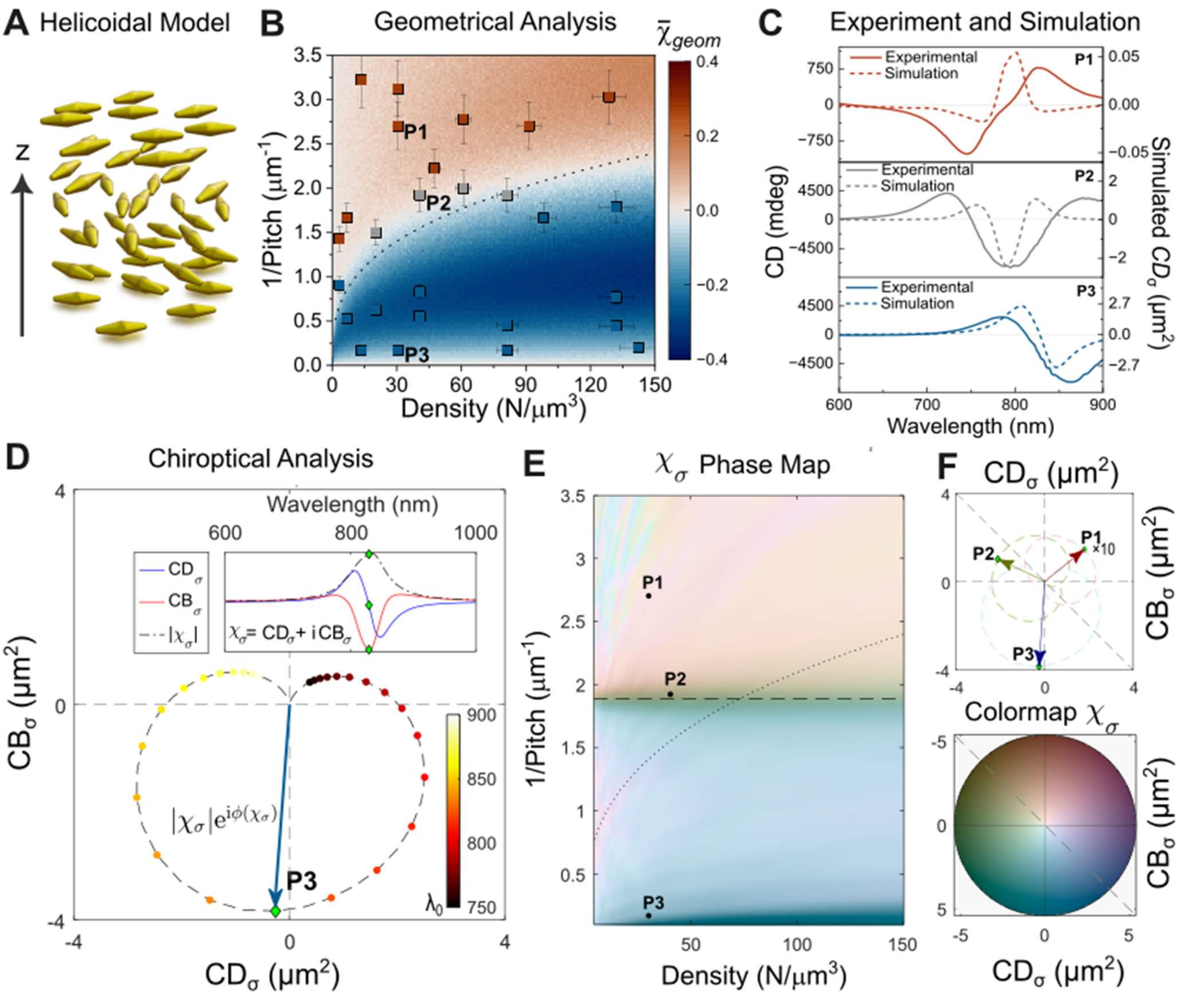

**Figure 3. Geometrical and chiroptical analysis.** (**A**) Schematic of the *Helicoidal Model.* (**B**) Mean geometrical chirality index ($\bar{\chi}_{geom}$) versus inverse pitch and particle density. The heatmap shows simulated left- and right-handed configurations in blue and red, respectively; square symbols show experimental circular dichroism (CD), color-coded by CD value. (**C**) Representative experimental (solid) and simulated (dashed) CD spectra for selected phase-diagram regions. Blue and red spectra show left- and right-handed bisignate signals; gray spectra show a left-handed Cotton effect. Each simulated spectrum is the average of 50 disorder realizations containing 1000 AuBPs. (**D**) Simulated chiroptical index for the *Helicoidal Model* for sample P3 (1000 AuBPs, density = 31 μm$^{-3}$, pitch = 5.9 μm, $n_{\text{Chitin}}$ = 1.5) using the T-matrix approach. The chiroptical index forms a limaçon-shaped trajectory in the complex plane. Green diamonds mark the plasmon resonance; the arrow indicates the magnitude and phase of the index at resonance. The color bar denotes clockwise wavelength progression. Inset: real, imaginary and magnitude spectra. (**E**) Phase map of the complex chiroptical index at the resonance ($\lambda_0 =$ 795 nm) versus density and pitch for one 1000-particle realization. The bivariate colormap encodes $|\chi_\sigma|$ (saturation), and $\phi$ ($\chi_\sigma$) (hue). (**F**) Simulated chiroptical indices for samples P1-P3.

## Chiroptical index and electromagnetic simulations

We developed an electromagnetic descriptor that could map the parameter space, analogous to that in Fig. 3B, while capturing both phase and magnitude of the chiroptical response. To this end, we defined a complex index that enables the screening of the electromagnetic response at a single wavelength. We combined CD and circular birefringence (CB) into a unified representation of electromagnetic chirality, $\chi_\sigma = CD_\sigma + \mathrm{i}\, CB_\sigma$ (*43*). Here, $CD_\sigma$ is the difference in extinction cross-section between left- and right-circularly polarized illumination, and $CB_\sigma$, its Kramers-Kronig pair, describes the circular birefringence cross-section. Within this framework, the phase of $\chi_\sigma$ defines a *chirality phase*, $\phi(\chi_\sigma)$, while its magnitude quantifies its strength, $|\chi_\sigma|$. Unlike conventional pseudoscalar chirality descriptors, the two-dimensional $\chi_\sigma$ avoids the false-zero paradox (*44*, *45*). Figure 3D shows the averaged $\chi_\sigma$ of 50 independent realizations generated using the *Helicoidal Model* for sample P3. An ensemble size of 1000 particles per realization is chosen as a compromise between computational tractability and representative sampling of the structure. The inset in Figure 3D denotes the real, imaginary, and magnitude spectra of $\chi_\sigma$. The complex-plane representation provides a visualization of the chiroptical response encoding its strength and phase. At plasmonic resonance, $\chi_\sigma$ traces a limaçon trajectory (here, a cardioid; fig. S18) in the complex plane. The apex (green diamond) corresponds to the resonance wavelength. The distance of the apex from the origin measures the maximum chiroptical strength, whereas its angular position defines the chirality phase. Importantly, the orientation of the limaçon determines the spectral shape of the chiroptical response, spanning bisignate and Cotton-type spectral line shapes (fig. S19). We first applied the complex-plane representation to the *Tomographic Dataset* (fig. S20). Tracing $\chi_\sigma$ at resonance reveals only small variations in the resonance wavelength across the subsets (fig. S20-S21) supporting their optical homogeneity. Moreover, replacing the experimentally measured particle orientations with ideal helicoidal orientations, or replacing the entire tomographic structure with the *Helicoidal Model* matched in pitch and density, produces only minor changes in the simulated spectra (figs. S22-S23). These results indicate close structural agreement between the experimental assembly and the *Helicoidal Model*. The low radiative ratio indicates that absorption, rather than scattering, dominates the chiroptical response (fig. S20). We therefore applied the electromagnetic framework to map the pitch-density design space using the *Helicoidal Model*. Using T-matrix simulations, we systematically varied the helicoidal pitch and particle density of ensembles containing 1,000 AuBPs arranged in a left-handed helicoidal structure. For each configuration, we evaluated the complex chiroptical index at the plasmonic resonance wavelength (795 nm). The resulting phase map (Fig. 3E) encodes the phase and magnitude of the chiroptical response through hue and saturation, respectively. The saturation reveals that the strongest chiroptical responses occur at larger pitches and higher particle densities, guiding our choice of parameters to experimentally achieve a high g-factor (Fig. 1C, fig. S24). The hue distinguishes different spectral responses, including positive and negative bisignate CD, as well as the negative Cotton-type line shape observed experimentally (Fig. 3C, P2). No positive Cotton-type response is found within the explored parameter space, suggesting that this state is inaccessible to the helicoidal architecture under the investigated conditions. Beyond these variations in strength and spectral line shape, the phase map reveals multiple inversions of chiroptical handedness within a template of fixed geometrical handedness. Two distinct inversion boundaries are identified, marked by the black dotted and dashed lines. The dotted boundary corresponds to the inversion predicted by the geometrical analysis (Fig. 3B), whereas the dashed boundary can be interpreted as a manifestation of a Borrmann-type effect (*46, 47*), arising from reduced absorption under left-handed circularly polarized illumination due to the periodic orientation of AuBPs along the local nematic director of the helicoidal architecture (fig. S25). Figure 3F illustrates this evolution for the three representative samples (P1-P3) with similar particle densities but different pitches. As the pitch decreases from P3 to P1, the resonance point rotates continuously through the complex plane, traversing three quadrants and ultimately inverting the chiroptical response despite the unchanged structural handedness of the template. The same pitch-dependent inversion is observed at a second particle density (fig. S26), demonstrating that this behavior is not restricted to a single density. Helical pitch and particle density therefore provide two structural parameters for controlling the handedness, line shape, and magnitude of the chiroptical response without changing the template.

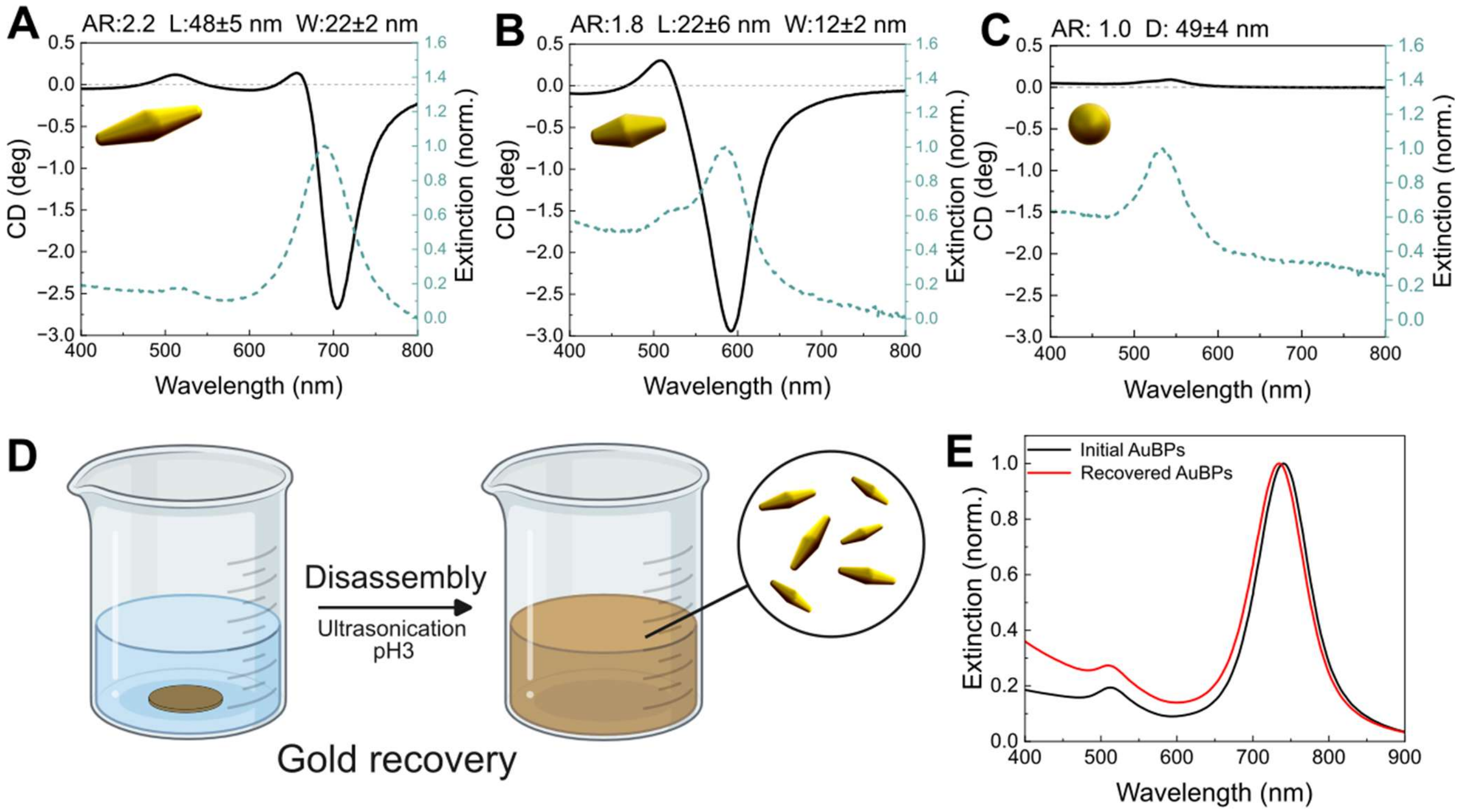

**Figure 4. Tunability and recovery of the gold particles.** (**A** to **C**) CD (solid curves) and normalized extinction (dashed curves) spectra of films containing gold particles with aspect ratios of 2.2 (A), 1.8 (**B**), and approximately 1.0 (**C**). Panel C shows nearly spherical particles. (**D**) Schematic of the film disassembly and gold-colloid recovery process. (**E**) Normalized extinction spectra of AuBPs before assembly and after recovery.

## Tunability and recovery of colloidal Au nanoparticles

The inclusion of plasmonic particles largely determines the chiral optical properties of the metamaterial. We therefore varied their aspect ratio to tune the position of the spectral response. The co-assembly process was tolerant to particles with different aspect ratios, allowing the optical properties of the material to be tuned while maintaining strong optical activity. As the particle absorption band shifted from 705 to 590 nm, the chiral plasmonic peak shifted accordingly (Fig. 4A, 4B). Interestingly, spherical particles produced a much weaker CD signal (Fig. 4C), confirming the importance of particle anisotropy for strong chiroptical activity. Finally, we tested the reversibility of the assembly process, allowing the metamaterial to be disassembled and its colloidal component recovered after use (Fig. 4D). We protonated the ChNC surface amines by lowering the pH, then used ultrasonication, filtration, and mild centrifugation to separate and recover the denser AuBPs from the ChNCs (see Materials and Methods).

The recovered AuBPs showed a largely unchanged plasmonic response, indicating that their anisotropic shape and colloidal stability were preserved (Fig. 4E). The recovery of the AuBP colloid supports material circularity and suggests that their incorporation is primarily driven by geometric confinement and organization within the helicoidal matrix.

## Conclusion

This work presents a scalable route to hierarchical metamaterials with strong plasmonic optical activity (g-factor > 0.1) and tunable handedness, achieved by co-assembling metallic particles with chitin nanocrystals. The spontaneous formation of a chiral nematic phase by the nanocrystals enables robust, reproducible fabrication without the stringent synthesis requirements of high g-factor plasmonic particles. Notably, the same left-handed template can produce either right- or left-handed chiroptical responses. Structural and optical characterization, combined with T-matrix simulations, allowed us to rationalize a structure-dependent mechanism of chirality transfer that may apply to other helicoidal systems. Given the diversity of anisotropic building blocks, including dielectric and emissive particles, this approach could enable functional materials with tailored chiral optical responses. Our findings show that chirality in hierarchical optical materials is a multiscale property and does not necessarily follow the handedness of the overall structure.

**Acknowledgments:** We thank Dr. Junyoung Kwon and Prof. Reiko Kuroda for help with the CD measurements, Dr. Yu Ogawa for helping with the electron microscopy characterizations, and Prof. Stephane Parola for providing the initial test sample of AuBPs. **Funding:** This work was made possible by the support of the Max Planck Society. T.J. was also supported by the NSERC Postdoctoral Fellowship (PDF-578177-2023). T.G.P. was also supported by the Marie Curie Postdoctoral Fellowship (101154876 – CINCOS). S.V. and B.F-P. acknowledge the EPSRC project “Bioderived and Bioinspired Advanced Materials for Sustainable Industries” EP/W031019/1 and the BiTe ERC-2020-CoS-101001637 (S.V. and B.F-P.). A.C. acknowledges the Codeχ - Starting grant (DFG). **Author contributions:** A.C. and S.V. conceived the project. A.C. prepared the samples and performed most of the experimental characterization. A.R. developed the analytical models and conducted most of the simulations. B.K. and T.J. assisted with the synthesis of ChNCs. E.R. and C.S. performed FIB-SEM measurements. A.C. and A.R. interpreted the data. T.G.P., B.F.-P., Z.L., and K.V. contributed to calculations and discussions on the configurational chirality. A.C. and A.R. prepared the final figures. A.C., A.R. and S.V. wrote the manuscript with comments from all co-authors.
**Competing interests:** No competing interests to declare.

**List of Supplementary Materials.**
Materials and Methods
Supplementary Text
Figs. S1 to S26
References *(1-58)*

# Supplementary Materials for

## Chirality Inversion and Giant Optical Activity in Scalable Self-Assembled Metamaterials

Antonio Carone[1†*], Aso Rahimzadegan[1†], Birgül Kalkan[1], Thomas G. Parton[1], Tony Jin[1], Zihao Lu[2], Clemens Schmitt[1], Emeline Raguin[1], Bruno Frka-Petesic[2,3], Kevin Vynck[4], Silvia Vignolini[1,2*].

[1] Max Planck Institute of Colloids and Interfaces (MPIKG), Potsdam, Germany.
[2] Department of Chemistry, University of Cambridge, Cambridge, UK.
[3] International Institute for Sustainability with Knotted Chiral Meta Matter (WPI-SKCM[2]),
1-3-1 Kagamiyama, Hiroshima University, Higashi-Hiroshima, Hiroshima, 739-8531, Japan.
[4] Institute of Light and Matter, Claude Bernard University Lyon 1, CNRS, Villeurbanne, France.

† These authors contributed equally to this work.
*Corresponding authors. Email: sv@mpikg.mpg.de, antonio.carone@mpikg.mpg.de

## 1 - Materials and methods

Shrimp shell chitin (*Pandalus borealis*) and potassium hydroxide (KOH) were purchased from Sigma-Aldrich. Hydrochloric acid (HCl, 37% and 1 M) and hydrogen peroxide ($H_2O_2$, 30% w/v) were obtained from Roth. Common white mushrooms (*Agaricus bisporus*) were sourced from a local supplier. Tetrachloroauric acid trihydrate ($HAuCl_4 \cdot 3H_2O$, 99.9%) was purchased from Roth, while silver nitrate ($AgNO_3$, 99.99%) and hexadecyltrimethylammonium bromide (CTAB, ≥99%) were purchased from Thermo Fisher Scientific. Hexadecyltrimethylammonium chloride (CTAC, 99%) was purchased from Acros Organics. 8-Hydroxyquinoline (99%) and sodium borohydride ($NaBH_4$, 98%) were purchased from Thermo Fisher Scientific. Catechol (98%) was purchased from TCI Chemicals, and citric acid monohydrate (99%) was obtained from MP Biomedicals. Ammonia solution ($NH_3$, 28%) was purchased from Alfa Aesar. Cellulose nitrate filters were purchased from Sartorius. Unless otherwise specified, Milli-Q water with a conductivity of 0.05 µS/cm was used throughout the experiments.

### Film fabrication

<u>Extraction and purification of chitin</u>

Shrimp-derived chitin was obtained from commercial powder and was further purified prior to use. In a typical purification process, chitin powder (10 g) was mixed with an HCl solution (300 mL, 0.01 M) and stirred at room temperature for 18 h to remove residual minerals. The powder was collected by vacuum filtration and transferred to a 500 mL round-bottom flask equipped with a condenser. Subsequently, a KOH solution (200 mL, 0.5 M) was added, and the mixture was stirred at 70 °C for 3 h to remove proteins. While maintaining the mixture at 70 °C under stirring, a concentrated solution of $H_2O_2$ (25 mL, 30% w/v) was slowly added dropwise. The $H_2O_2$ addition was repeated in total three times with a 1 h interval between different additions for a total volume added of 75 mL. Caution must be taken because adding $H_2O_2$ too quickly could cause excessive foaming. The mixture was then diluted with water, the powder was collected by vacuum filtration, and washed with water. The entire purification process was then repeated. The aspect of the final product was a white, fine powder stored at room temperature and constituted the starting purified shrimp chitin.

Fungal-derived chitin was isolated from the fruiting bodies of common white mushrooms. A total of 2 kg of mushrooms were first rinsed with warm water, then boiled in water for 1 h. Boiled mushrooms were blended in a household blender to obtain a homogeneous purée, which was boiled again and collected by filtering and pressing through 100 µm mesh nylon cloth, yielding a brownish-gray pulp. The obtained pulp was subjected to base treatment followed by a basic bleaching. For the base treatment, the mushroom wet pulp (≈ 40% w/w solid content, 40 g) was dispersed in a KOH solution (400 mL, 0.5 M) at 70 °C in a 1-L round-bottom flask equipped with a condenser and stirred for 3 h, and the mixture was quenched by dilution with 400 mL of cold Milli-Q water. Solids were collected by vacuum filtration and washed extensively, and the base treatment was repeated. After the base treatment, the mushroom pulp was subjected to a basic bleaching. For the basic bleaching, the pulp was stirred in a KOH solution (400 mL, 0.5 M) at 70 °C, and a concentrated solution of $H_2O_2$ (75 mL, 30% w/v) was added slowly during stirring to prevent foaming while the reaction occurred. The reaction mixture was then quenched after 3 h with 400 mL of cold Milli-Q water. The solids were collected with vacuum filtration and washed extensively until reaching neutral pH. Then, the base treatment and basic bleaching were repeated. The resulting wet sample was stored at 4 °C and constituted the starting purified fungal chitin.

Preparation of chitin nanocrystals (ChNCs)

The ChNCs were prepared according to a protocol adapted from Narkevicius *et al.* (*42*) and consisted in performing acid hydrolysis of previously purified chitin powder.

To produce fungal or shrimp ChNCs, the purified chitin was subjected to acid hydrolysis. A total of 10 g of chitin was added into 300 mL of a 3 M HCl solution preheated to reflux under vigorous stirring. After 270 min (shrimp) or 180 min (fungal), the reaction was quenched by pouring the mixture into ice-cold water. The solid fraction of the mixture was washed twice by centrifugation (4000 RCF, 5 min) and redispersed in water by vigorous shaking.

A final centrifugation and redispersion step were performed at a higher centrifugal force (12,000 RCF, 30 min). The mixture was then dialyzed against distilled water until the conductivity of the dialysis bath remained stable overnight. The final dialysis steps were performed against Milli-Q water.

The mixture was then diluted with water to reach approximately 400 mL, and the pH was adjusted to 4 by adding HCl. The resulting ChNC suspension was tip-sonicated using an ultrasonic probe (Sonic Vibracell 500 W, 30% amplitude, tip diameter = 12.7 mm) in pulse mode (10 s on, 15 s off) for a total of 540 s on. The suspension was then filtered through 8.0 µm and 0.8 µm filters (Cellulose Nitrate Filter, Sartorius), and the filtered suspension was collected. Transmission electron microscopy (TEM) images of the produced ChNCs are shown in figure S1.

Synthesis of elongated gold bipyramids (AuBPs) with different aspect ratios

AuBPs with different aspect ratios (1, 1.8, 2.2, 3.2) were prepared according to a protocol fully described by Chateau *et al.* (*48*). TEM images of the prepared particles are shown in figure S2.

Self-assembly of ChNCs into a chiral nematic phase

Self-assembly of ChNCs was achieved by adjusting the HCl content of the suspension (*31*). The chitin concentration was first determined by thermogravimetric analysis, and the suspension was then dialyzed against HCl to reach a final ratio of 100 mmol HCl per kg of ChNCs. After dialysis, the suspension was transferred into a 250 mL PTFE flask and concentrated using a rotary evaporator to roughly 8 wt% (shrimp) and 5 wt% (fungal), and stored in a fridge at 4 °C.

The self-assembly of the ChNCs suspension was studied by diluting the stock suspension to achieve the desired concentration in a vial and leaving it undisturbed until a clear phase separation was observed. A concentration of 6.15% wt (shrimp) and 3.25% wt (fungal) was used to induce a macro-phase separation. Subsequently, the chiral nematic anisotropic phase (a-phase) and the upper isotropic phase (i-phase) were separated with a pipette.
Encapsulation of the solution in capillaries was done by introducing a few μL into a glass capillary and sealing it on both ends with a hot glue gun. The capillaries were allowed to stand vertically for a week until the suspensions reached equilibrium.

Tuning of the pitch.
Pitch tuning was achieved by mixing the a-phases of previously prepared fungal and shrimp ChNC suspensions at the desired dry-weight ratios (fig. S16). The relative amounts of each suspension were calculated from their respective ChNC concentrations, so that the reported fungal/shrimp ratios refer to the mass fraction of ChNCs rather than the volume of suspension mixed.

Co-assembly of AuBPs with ChNCs and chiral plasmonic films.
In a typical synthesis, the a-phase of ChNCs was mixed with a concentrated AuBPs solution in a ratio of 50/1 (ChNCs/AuBPs, v/v). This ratio can be modified; however, excessive dilution of the a-phase should be avoided, as it can phase-separate into a biphasic (isotropic-anisotropic) suspension.
The concentration of the AuBP suspension was adjusted by centrifugation of the stock solution ([$Au^0$] ≈ 5 g/L, 5 mM CTAB). To ensure the colloidal stability of the AuBPs, the CTAB concentration in the final dispersion was adjusted to 1 mM. After mixing, the suspension was homogenized by vortexing for 1 min, cast into the desired mold, and allowed to evaporate slowly. Typically, a 1 mL suspension in a plastic Petri dish loosely covered with a laboratory paper towel dries within 1-2 days.
Alternatively, the self-assembly and the drying step can be temporally separated. After casting, the mixture was allowed to rest for 12 h on top of a commercial neodymium magnet (NdFeB, 40 × 40 × 20 $mm^3$, grade N42). Magnetic fields can play a role in aligning chiral nematic phases of ChNCs, similar to what is observed for cellulose nanocrystals *(43, 49, 50)*.
Parafilm was used to prevent evaporation. After 12 h, the film can be physically cross-linked by exposure to $NH_3$ gas. The consequent increase in pH causes the deprotonation of surface amines on the ChNC suspension. Because the process is governed by diffusion, the time required for complete gelation depends on the thickness of the suspension. For a 1 mL suspension in a plastic Petri dish, gelation can occur within 10 min. The film is then left to dry. These additional steps can promote the formation of long-range order in the cholesteric domain (fig. S3). This procedure, involving gelation with ammonia gas prior to drying, can also be employed to produce achiral plasmonic films by mixing AuBPs with the i-phase of ChNCs. Films cast in a plastic Petri dish can be easily detached upon drying, yielding self-standing films. For films containing high AuBP concentrations, the film thickness was reduced to maintain sufficient light transmission. This was achieved by reducing the casting volume. In these cases, around 50 μL of suspension was cast onto a glass substrate using a circular polydimethylsiloxane (PDMS) mold with a diameter of 1 cm.

Gold colloidal suspension recovery
In order to recover the gold as a stable colloidal suspension, the dried film was immersed in a solution of HCl (pH 3) containing CTAB (5 mM) for 1 h to promote swelling. The resulting suspension was then tip-sonicated (30 min, pulse 10/15 s, using a 12.7 mm tip). The suspension was subsequently filtered using 8.0 µm and 0.8 µm cellulose nitrate filters, and the solution was finally washed by centrifugation (5,000 RCF for 10 min) three times with a solution of HCl (pH 5) containing CTAB (5 mM) to separate the AuBPs from the ChNCs.

**Characterizations**

Circular dichroism spectroscopy

Circular dichroism (CD) spectra were initially collected using a Chirascan qCD spectrometer (Applied Photophysics). For film measurements, the samples were mounted on a rotating holder and positioned with the incident beam normal to the sample surface. Circular dichroism is defined as the differential absorption of left- and right-circularly polarized light:

$$\mathrm{CD} = \Delta A = (A_L - A_R) \tag{1}$$

where $A_L$ and $A_R$ are the absorbances measured under left- and right-circularly polarized (LCP and RCP) illumination, respectively. In commercial CD spectrometers, the CD signal is generally reported as ellipticity. Under the assumption that the differential absorption is small compared with the overall absorption, the ellipticity in degrees is related to the differential absorbance by

$$\mathrm{CD}_{\mathrm{deg}} \approx 32.982 \times \Delta A \tag{2}$$

A dimensionless dissymmetry factor (g-factor) is therefore commonly used to quantify chiroptical efficiency relative to the overall extinction. Generally, for samples that both absorb and scatter light, such as plasmonic nanostructures, extinction accounts for both absorption and scattering. Accordingly, the g-factor can be expressed as

$$\mathrm{g-factor} = \mathrm{CD}_{\mathrm{deg}} \,/\, (32.982 \times A). \tag{3}$$

where $A$ indicates the extinction of unpolarized light. Extinction and CD spectra were therefore recorded simultaneously under identical measurement conditions.

In addition, anisotropic solid films can exhibit linear dichroism (LD) and linear birefringence (LB), which may contribute to an apparent CD signal. The films were measured at four different orientations: at orthogonal azimuthal rotations (0°, 90°) around the axis normal to the surface and after flipping the sample (i.e., reversing the front/back orientation). A genuine CD contribution is invariant upon these operations, whereas contributions associated with LD and LB exhibit systematic changes (*32-34*). The final CD spectra were obtained by averaging the measurements acquired at the different orientations to minimize residual contributions from linear anisotropy.

Circular dichroism measurements using a custom optical setup.

The accurate measurement of large CD signals (above 2000 mdeg), as in our case, presents another challenge for conventional CD spectrometers. As previously mentioned, commercial instruments generally rely on the approximation that the CD is small compared to the overall absorbance (*51*). These approximations from the measured polarization modulation can introduce errors when the dichroic response is large. To overcome these limitations, we directly measured and compared the absorbance of left- and right-circularly polarized light separately using a custom optical setup.

Broadband illumination was provided by a xenon lamp (Ocean Optics HPX2000) coupled to a reflective collimator (Thorlabs RC08SMA-F01) via an optical fiber (200 µm core diameter, Avantes FC-UV200-2-SR). A linear polarizer (Thorlabs WP25M-UB) and rotatable superachromatic quarter-wave plate (Halle RSU 1.4.15, B.) were placed in the beam path before the sample to provide, upon rotation of the quarter-wave plate, either LCP or RCP illumination. The transmitted light was collected by a second reflective collimator and coupled into a spectrometer (Avantes AvaSpec-HS2048) through a second optical fiber (600 µm core diameter, Avantes FC-UV600-2-SR). In this case, the g-factor was calculated from the measured difference in transmission with $A = -log_{10}(\frac{\mathrm{I}}{\mathrm{I_0}})$.

$$\mathrm{g-factor} = \frac{2(A_\mathrm{L} - A_\mathrm{R})}{A_\mathrm{L} + A_\mathrm{R}}. \quad (4)$$

The CD values obtained from the direct extinction measurements were converted to ellipticity in degrees to express the chiroptical response on the same scale as conventional CD measurements and enable direct comparison across the different datasets. The conversion was performed using the exact expression, rather than the small-signal approximation:

$$\mathrm{CD_{deg}} = \frac{180}{\pi} \arctan\left[\tanh\left[\frac{\ln(10)\Delta A}{4}\right]\right] \quad (5)$$

<u>Spatially resolved circular dichroism measurements</u>

The films exhibiting the highest chiroptical response also contained the highest density of particles. This resulted in a spatial heterogeneity in the distribution of the gold nanoparticles. In particular, regions with locally increased nanoparticle density could form particle-rich domains that do not let light through (fig. S7). When measured over a large area, these contributions increase the total extinction while contributing little to the differential response, thereby diluting the area-averaged g-factor. To distinguish the local chiroptical response from this spatial averaging effect, CD measurements were therefore performed using a microscope-based optical setup. The localized illumination and collection geometry enabled measurements from selected regions of the films, allowing the spatial variation of the chiroptical response and the local g-factor to be resolved.

Spatially resolved optical measurements were performed using a Zeiss Axioscope 7 microscope operated in transmission mode with a 10× objective (Zeiss EC Epiplan-Neofluar, NA = 0.3). To measure the transmission under different circular polarization states, the incident light was passed through a linear polarizer (Thorlabs WP25M-UB) followed by a rotatable superachromatic quarter-wave plate (Halle RSU 1.4.15, B.), generating either left- or right-circularly polarized (LCP or RCP) illumination upon rotation of the quarter-wave plate. The polarized light was focused onto the sample through the microscope condenser. Transmitted light was collected by the microscope objective and directed by a beamsplitter either to a CMOS camera (IDS UI-3580LE-C-HQ) for imaging or to an optical fiber (600 µm core diameter, FC-UV600-2-SR) coupled to a spectrometer (Avantes AvaSpec-HS2048). The optical fiber was positioned conjugate to the microscope image plane, enabling spatially resolved spectra. For each measurement position, transmission spectra were acquired under both LCP and RCP illumination and converted to extinction. The local CD and g-factor were then calculated from the corresponding LCP and RCP extinction spectra. Measurements were performed at multiple positions across the films to assess the spatial variation of the chiroptical response.

Scanning electron microscopy (SEM)

SEM characterization was performed using either a Tescan CLARA or a Thermo Fisher Scientific E-SEM Quattro, depending on instrument availability. Cross-sectional images were acquired using a 90° specimen mount (10 mm diameter) with an accelerating voltage of 3-5 kV and a working distance of 5-10 mm. Images of the helicoidal ChNC structure were acquired using an Everhart-Thornley detector, whereas AuBPs were visualized using an annular backscattered electron detector. The pitch of the dried films was measured from cross-sectional SEM images of the helicoidal ChNC architecture. Cross-sections were prepared by manually fracturing the dried films by gently pulling them from opposite sides. The pitch was determined as twice the periodicity observed in the cross-section, corresponding to a full 360° rotation of the helicoidal structure. Pitch values were obtained from multiple regions of each film and reported as the mean ± standard deviation. Prior to imaging, samples were sputter-coated with either platinum or carbon.

Transmission electron microscopy (TEM)

TEM images were acquired using a JEOL JEM-F200 operating at 200 kV and a TVIPS TemCam-F216 camera. ChNC samples were prepared by casting a suspension of 0.001 wt% ChNC, adjusted to pH 3 with diluted HCl, onto a carbon-coated copper grid prepared by glow discharge and stained with uranyl acetate solution. AuBP samples were prepared by casting a suspension at 1 mM CTAB onto a carbon-coated copper grid. The micrographs were analyzed using Fiji software.

Focused Ion Beam- Scanning Electron Microscopy (FIB-SEM) and 3D reconstruction.

The samples were mounted on a 1 cm diameter stub using conductive carbon tape and subsequently coated with 15 nm carbon and 5 nm of platinum to improve conductivity and reduce charging and beam damage. FIB-SEM slice-and-view imaging was performed using a Zeiss Crossbeam 540 (Zeiss Microscopy GmbH, Oberkochen, Germany). The exposed surface was polished using a low ion beam current (700 pA) and imaged with the electron beam at an accelerating voltage of 2 kV and a beam current of 150 pA. Images were acquired using a combination of secondary electron (SE) and backscattered electron (ESB) detectors to enhance contrast. Sequential images were collected using a slice-and-view protocol with all voxel dimensions at 8 nm and an image resolution of 3072 × 2304 pixels. Stack alignment and segmentation were performed using Dragonfly software (Object Research Systems (ORS) Inc., Montreal, Canada). The final region of interest was cropped to a volume of 16 × 10 × 10 $\mu m^3$. The position of each AuBP, defined as the center of mass, and orientation, defined as the principal long axis of each particle, were extracted using Dragonfly software. The final tomographic dataset shape being a parallelepiped and not a rectangle was caused by instrumental drift, which caused the milled region to shift upward during the FIB process.

Calculation of the density of particles

The particle density obtained from the tomographic dataset was used as a reference, and the particle densities corresponding to different amounts of added AuBPs were estimated accordingly.

In situ monitoring of the self-assembly.

The pitch of the suspensions was monitored by preparing glass capillaries sealed with a glue gun or 1 mm path-length glass cuvettes sealed with Parafilm, which were left to equilibrate for at least one week. The characteristic fingerprint texture of the chiral nematic phase appeared as alternating dark and bright stripes when viewed between crossed polarizers (figs. S1 and S4). It is important to note that ChNCs are birefringent only when immersed in a medium with a different refractive index, like water (extrinsic birefringence). The pitch of the anisotropic phase was determined as twice the periodicity of the fingerprint texture observed in optical microscopy images. Images were acquired in transmission mode between crossed polarizers using a Keyence VHX-7100 microscope.

Helicoidal Model

As full tomographic data are not available for all configurations, we developed an empirically informed model guided by the key features of the system. The bipyramid orientations follow the helicoidal template, while the particle positions are reproduced using a tailored point process that matches the empirical spatial statistics. Together, these elements enable a model for simulating the optical response across a range of particle densities and helicoidal pitches, using the newly introduced metrics.

Tailored Disorder Point Process

To reproduce the pair correlation function and nearest-neighbor statistics of the *Tomographic Datasets*, we constructed a tailored disordered point process in three dimensions. Starting from a Poisson configuration, points are sequentially removed according to a distance-dependent rule defined by a hard-core exclusion radius and a soft-core shell, and are iteratively reinserted to recover the target density. The process is inspired by hard-core Matérn processes and Gibbs-type interactions (*52*). Details of the algorithm are shown in Figs. S12-S13.

Evaluation of the $S_2$ parameter

The orientational order of the AuBPs in the *Tomographic Datasets* was quantified by calculating the nematic order parameter $S_2$. For each particle j, the orientation was described by a unit vector $\hat{\mathrm{x}}_{\mathrm{j}}$ corresponding to the long axis of the particle. The $S_2$ parameter was obtained as the largest eigenvalue of the corresponding tensor, with α and $\beta$ being the vector components along the *x, y,* and *z* axes, and $\delta_{\alpha\beta}$ the Kronecker delta.

$$S_{\alpha\beta} = \frac{1}{N}\sum_{j=1}^{N}\left[\frac{3}{2}\hat{\mathbf{x}}_{\mathrm{j}\alpha}\hat{\mathbf{x}}_{\mathrm{j}\beta} - \frac{1}{2}\delta_{\alpha\beta}\right] \tag{6}$$

Because the helicoidal structure contains a continuously rotating director, calculating $S_2$ over the entire volume would average out the local orientational order. We therefore calculated $S_2$ within thin *z*-slices of approximately 100 nm thickness, over which the director orientation can be considered locally uniform. The robustness of the $S_2$ analysis was assessed by slightly varying the slice thickness. Across the tested range, the calculated values remained between approximately 0.7 and 0.8.

Quantification of geometrical chirality

The geometrical chirality of the AuBP assemblies was evaluated using a pseudoscalar chirality index defined as

$$\chi_{\mathrm{geom}} = (\hat{r}_{12} \cdot \boldsymbol{w_1} \times \boldsymbol{w_2})\,(\boldsymbol{w_1} \cdot \boldsymbol{w_2}) \tag{7}$$

where $\boldsymbol{w_1}$ and $\boldsymbol{w_2}$ are the orientation vectors of two neighboring particles and $\hat{r}_{12}$ is the unit vector connecting them (fig. S15A). This geometrical measure of chirality takes into account the head-tail symmetry of the AuBPs and gives values from -0.5 (left-handed) to 0.5 (right-handed), reaching its maximum magnitude for a dihedral angle of 45°. The use of a normalized vector prevents the interparticle distance from affecting the magnitude of $\chi_{\mathrm{geom}}$.
The average value of geometrical chirality $\bar{\chi}_{geom}$ was calculated to evaluate the net chirality arising from different configurations of pitch and particle density ($N = 1.3 \times 10^5$) generated with the *Helicoidal Model* (fig. S15B). Nearest neighbor pairs were identified using a k-dimensional tree search. To focus on pairs probing the helicoidal rotation, only pairs separated along the *z*-axis by more than 60 nm (roughly double the width of two particles) were included.

Dihedral angles and pitch inversion

For an isotropic particle distribution with number density $\rho$, the average nearest-neighbor distance is approximated by (*53*)

$$d = 0.55396/\rho^{1/3} \tag{8}$$

In an ideal left-handed helicoidal structure, the azimuthal angle difference, or dihedral angle, between two neighboring bipyramids separated by a distance *d* can be expressed as

$$\Delta\varphi = \frac{2\pi d}{P} \tag{9}$$

where *P* is the helical pitch.
For geometrical considerations, a handedness inversion is expected when the dihedral angle satisfies

$$\Delta\varphi = n\frac{\pi}{2} \tag{10}$$

where $n = 1, 2, 3\ldots$ denotes the possibility that this condition is met at multiple angle differences. Combining the dihedral angle relation with the inversion condition at $n = 1$ gives the pitch at which the inversion is expected $P_i$

$$2\pi\frac{d}{P_i} = n\frac{\pi}{2} \tag{11}$$

Solving for $P_i$ yields

$$P_i = 4d \tag{12}$$

Substituting Eq. 8 for $d$ gives

$$P_i = \frac{4 \times 0.55396}{\rho^{1/3}} = \frac{2.2158}{\rho^{1/3}} \tag{13}$$

Although the mean nearest-neighbor distance is estimated assuming an isotropic spatial distribution, the relevant separation for the helicoidal rotation is its axial component, $|\Delta z|$. However, the chiroptical contribution also depends on pair geometry: pairs stacked along the helical axis are expected to contribute more strongly than pairs lying within the same xy plane. The provided mean distance therefore is a heuristic length scale for estimating the inversion boundary.

Eq. 12 shares similarity with the Nyquist-Shannon sampling theorem, a foundational concept in signal processing and telecommunications (*54*). According to the theorem, a periodic function can be correctly reconstructed only if sampled at least twice its frequency. If this condition, referred to as the Nyquist-Shannon condition, is not satisfied, a different function is reconstructed (aliasing). Note that this theorem usually involves a factor of 2 (twice its frequency), while in our case the factor of 4 in Eq. 12 derives from how the periodicity is defined in the helicoidal structure (a full turn, or 360° rotation). The Nyquist-Shannon condition can be linked with chirality, considering our left-handed helix as a three-dimensional periodic function and the gold particles as the sampling. The periodicity and handedness of the chitin pitch can be seen as reconstructed locally by the AuBPs, in analogy to how a periodic function is reconstructed by sampling.

**Optical Simulations**

Chiroptical index

The extinction cross section for an electric dipolar particle is (55):

$$\sigma_{\text{ext}}(\omega) = -\frac{k}{\varepsilon}\,\text{Im}\big(\mathbf{E}(\omega).\,\mathbf{p}^{*}(\omega)\big) \tag{14}$$

where, $\mathbf{E}$ is the incoming electric field and $\mathbf{p}$, here, is the effective dipole moment defined as:

$$\mathbf{p}(\omega) = \frac{6\pi\varepsilon}{k^{3}}\boldsymbol{\alpha}(\omega)\mathbf{E}(\omega) \tag{15}$$

where $\boldsymbol{\alpha}(\omega)$ is the effective polarizability derived from the T-matrix approach (*see below*).
The circular dichroism cross section $\text{CD}_{\sigma}$ is defined as:

$$\text{CD}_{\sigma}(\omega) = \sigma_{\text{ext}}^{\text{LCP}}(\omega) - \sigma_{\text{ext}}^{\text{RCP}}(\omega) \tag{16}$$

The Kramers-Kronig pair of the $\text{CD}_{\sigma}$, i.e., the circular birefringence cross section $\text{CB}_{\sigma}$ can readily be accessed via

$$\text{CB}_{\sigma}(\omega) = \frac{k}{\varepsilon}\text{Re}[\mathbf{E}^{\text{LCP}}(\omega).\,\mathbf{p}^{*,\text{LCP}}(\omega) - \mathbf{E}^{\text{RCP}}(\omega).\,\mathbf{p}^{*,\text{RCP}}(\omega)\,] \tag{17}$$

Therefore**,** we define the chiroptical index ($\chi$) as a complex measure of electromagnetic chirality encompassing both the $\text{CD}_{\sigma}$ and $\text{CB}_{\sigma}$:

$$\chi_{\sigma}(\omega) = \text{CD}_{\sigma}(\omega) + \text{i}\text{CB}_{\sigma}(\omega) = \frac{\text{i}k}{\varepsilon}[\mathbf{E}^{\text{LCP}}(\omega).\,\mathbf{p}^{*,\text{LCP}}(\omega) - \mathbf{E}^{\text{RCP}}(\omega).\,\mathbf{p}^{*,\text{RCP}}(\omega)] \tag{18}$$

Transition matrix (T-Matrix) approach

To model the optical response of large ensembles of AuBPs, we employed the T-matrix formalism, which provides an efficient description of multiple scattering between arbitrarily positioned and oriented particles (56). In this framework, the electromagnetic field scattered by each particle is expanded into vector spherical wave functions, and the T-matrix relates the incident and scattered multipole coefficients. Once the T-matrix of an individual AuBP is known/simulated, the response of large particle assemblies can be calculated by accounting for all orders of electromagnetic coupling between the particles. Detailed derivations, notation conventions, and normalization schemes adopted throughout this work follow established formulations of the T-matrix method and are summarized in Refs. (*55*, *57*). An open-source T-matrix framework is accessible via TREAMS (*36*).

T-matrix calculation of an individual bipyramid

Transmission electron microscopy (TEM) reveals that the gold nanoparticles (AuBPs) adopt a curved bipyramidal morphology. Based on these images, the AuBPs are modeled as bicones with spherical tips (*58*) and their individual electromagnetic response (i.e., the polarizability/T-matrix) is computed using multipole excitation within a T-matrix framework code available in the GitHub repository (*36*) in COMSOL Multiphysics (fig. S17). We have used an electric dipole as sufficient for the simulation of the individual AuBP.

T-matrix simulation of the Tomographic Dataset

*Tomographic Dataset* provides both the spatial coordinates and orientations of the bipyramids. The *Dataset*, containing approximately $4 \times 10^5$ particles, is too large for direct full-scale simulation. Therefore, the *Tomographic Dataset* is partitioned into smaller sub-volumes, each supplemented with a buffer region to mitigate edge effects (fig. S21). A total of 510 sub-volumes are simulated using the T-matrix approach. The resulting dissymmetry g-factors are shown in Fig. 1C in the main text. The distribution of the chiroptical index $\chi$ across sub-volumes indicates a relatively homogeneous and well-dispersed nanoparticle assembly (fig. S20). In all the simulations, the chitin nanocrystal matrix is treated as an effective homogeneous medium with a refractive index of 1.5. For comparison, we performed additional simulations assuming idealized particle orientations ($S_2 = 1$) following a perfect helical pitch without out-of-plane deviations (fig. S22). The close agreement between the helicoidal model and the idealized case with the tomographic reconstruction indicates that the experimental structure already approximates a highly ordered configuration (fig. S23).

## 2 - Supplementary Figures

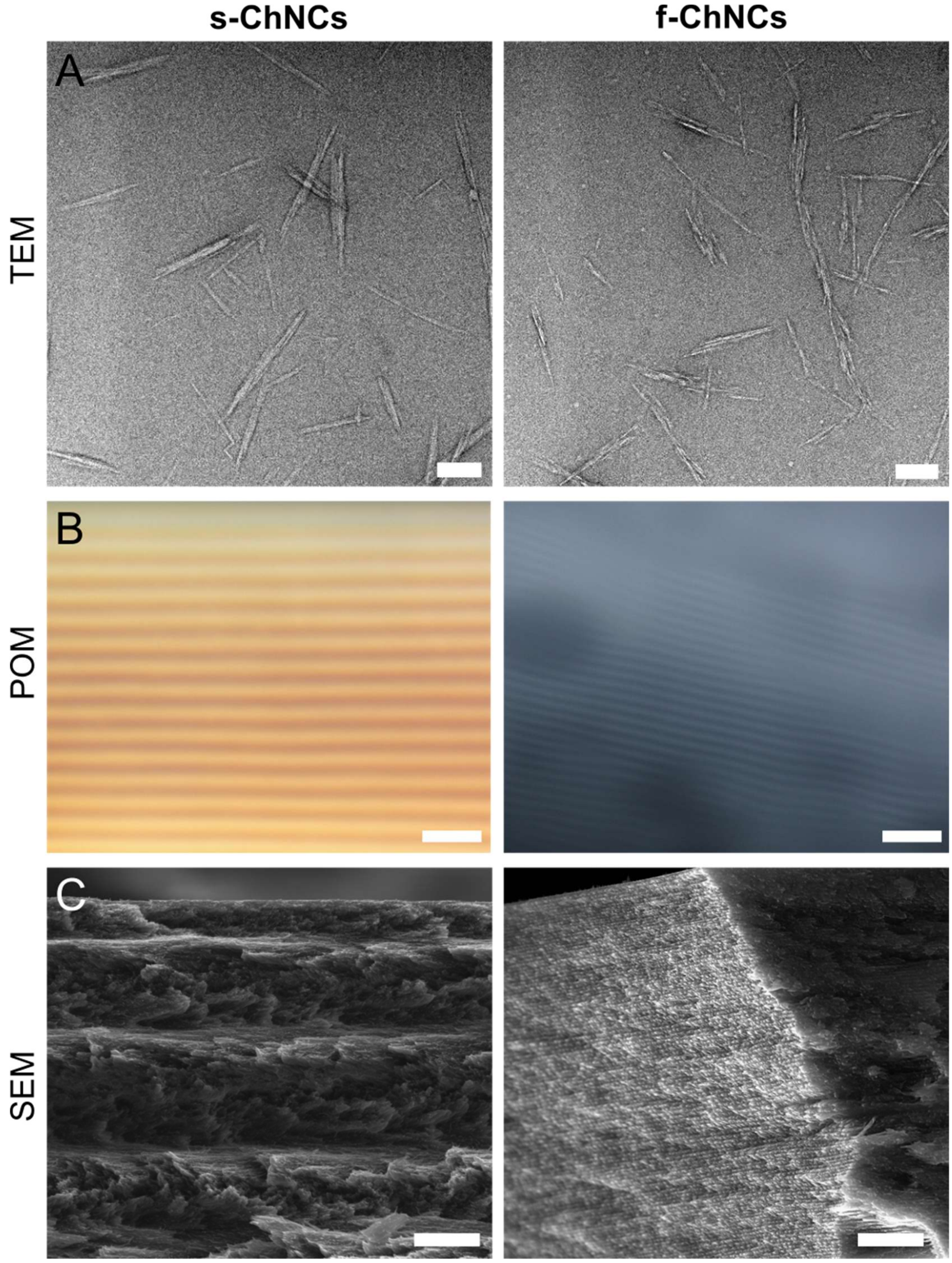

**Figure S1.** (A) TEM images of isolated *(left)* shrimp-derived chitin nanocrystals and *(right)* fungal-derived nanocrystals. Scale bars, 100 nm. (B) Polarized optical microscopy (POM) of capillaries showing the characteristic fingerprint of the chiral nematic phase obtained by self-assembly. Scale bars, 50 µm. (C) SEM of a cross-section of films showing the helicoidal structure with different periodicities between shrimp and fungal chitin. Scale bars, 2 µm.

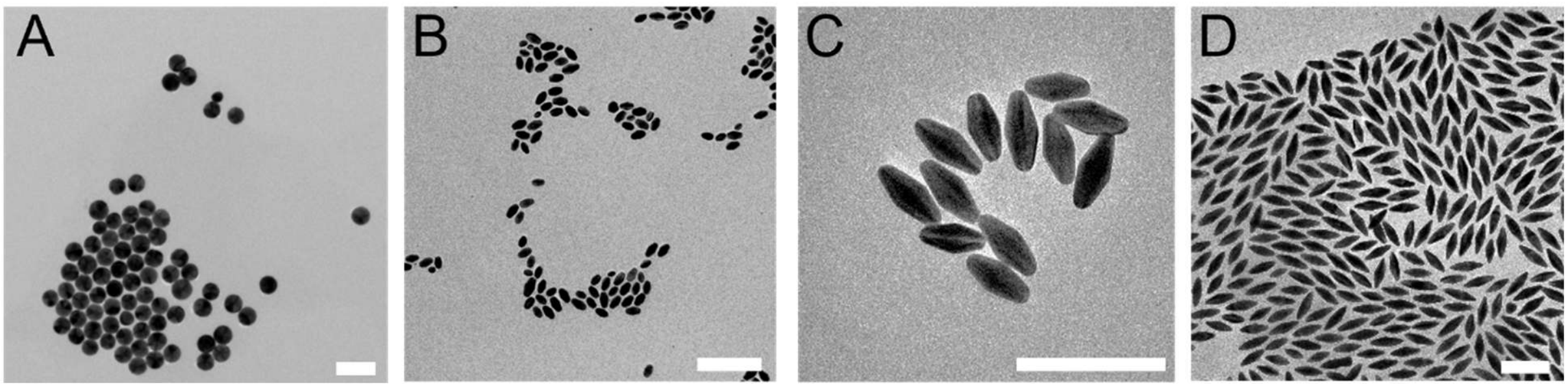

**Figure S2**. TEM images of the synthesized gold particles with different aspect ratios. (**A**) $diameter_{max}$: 50±4 nm, $diameter_{min}$: 48±4 nm, $diameter_{avg}$: 49±4 nm, aspect ratio 1.04. (**B**) Length: 22±6 nm, width: 12±2 nm, aspect ratio 1.8. (**C**) Length: 48±5 nm, width: 22±2 nm, aspect ratio 2.2. (**D**) Length: 68±19 nm, width: 21±3 nm, aspect ratio 3.2. The different scale bars are all 100 nm. Values are means ± SD from 100 particles.

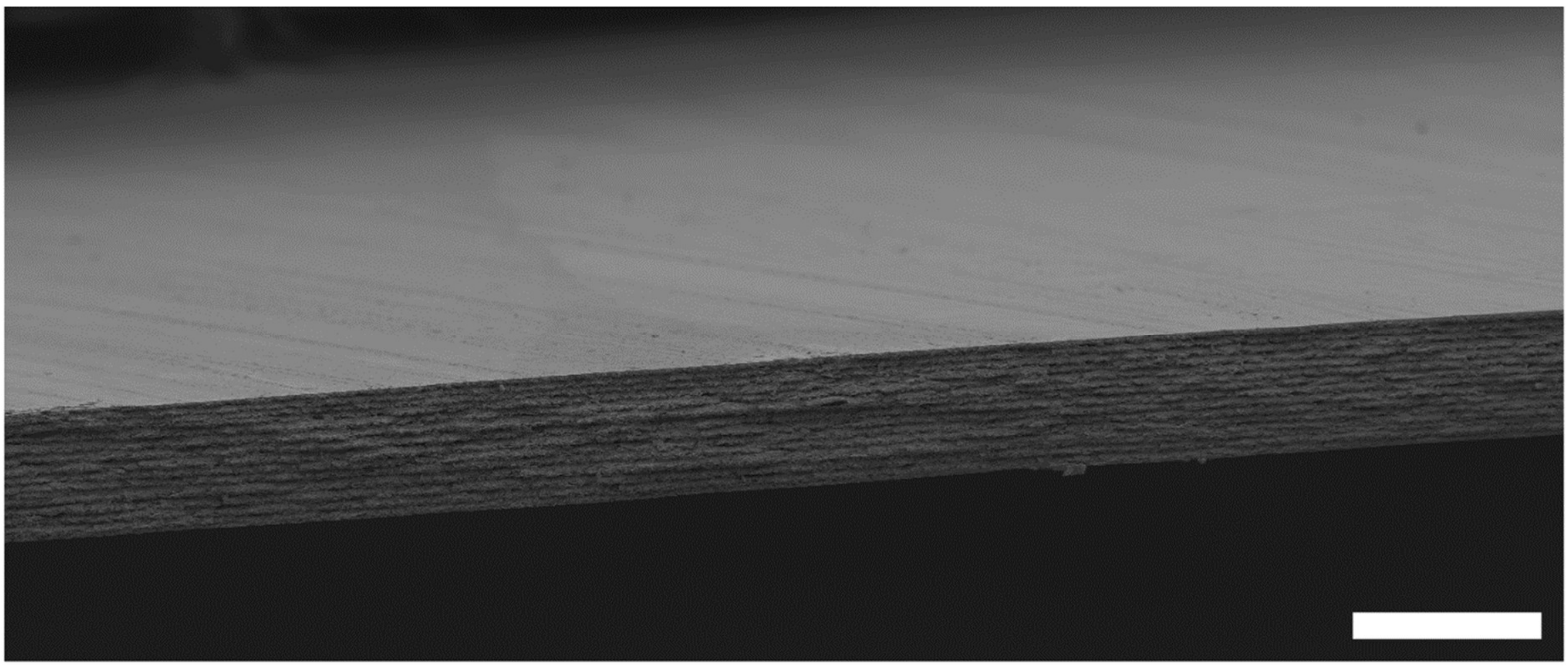

**Figure S3**. SEM cross-section of a film showing long-range order made by aligning the chiral nematic phase with magnets before drying. Scale bar, 50 μm.

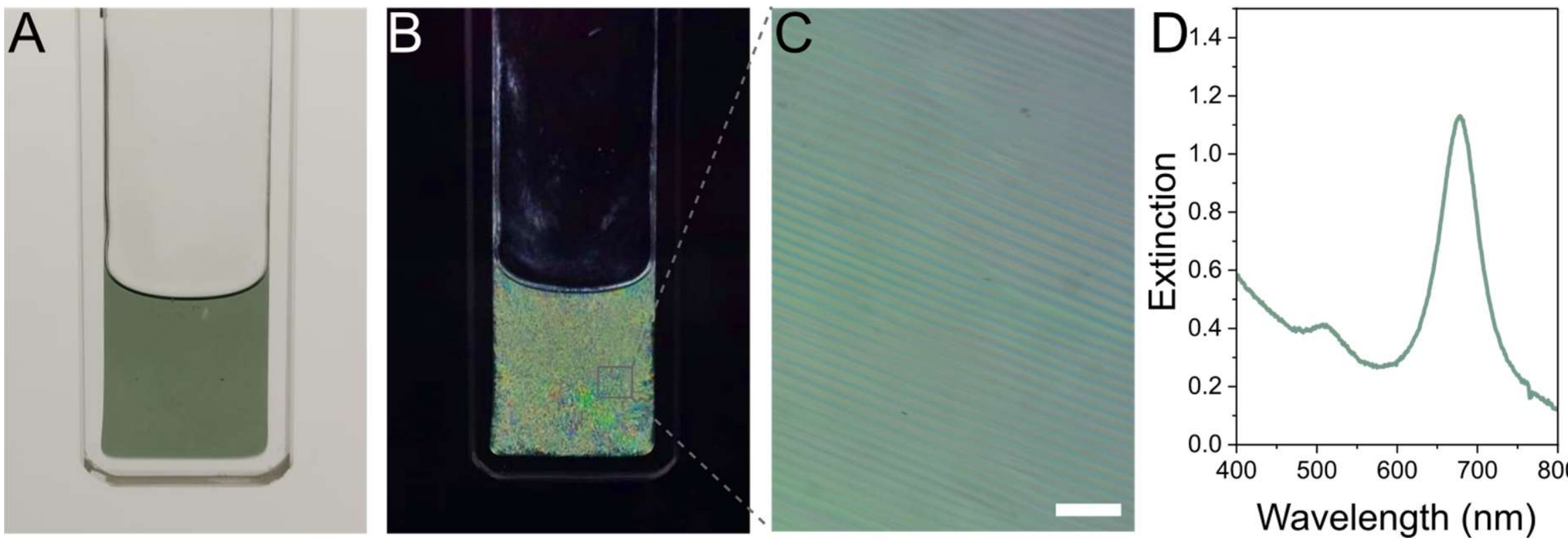

**Figure S4.** (**A**) Picture of a cuvette containing chitin nanocrystals and gold nanoparticles (aspect ratio 2.2) showing homogeneous green coloration caused by the presence of elongated plasmonic particles. (**B**) Picture of the cuvette between crossed polarizers showing the extrinsic birefringence of the chitin nanocrystals dispersed in water. (**C**) Polarized optical microscopy of the chiral nematic phase showing the formation of the characteristic fingerprint pattern in the presence of gold nanoparticles. Scale bar 250 μm. Overlay not to scale. (**D**) Relative extinction spectrum of the sample.

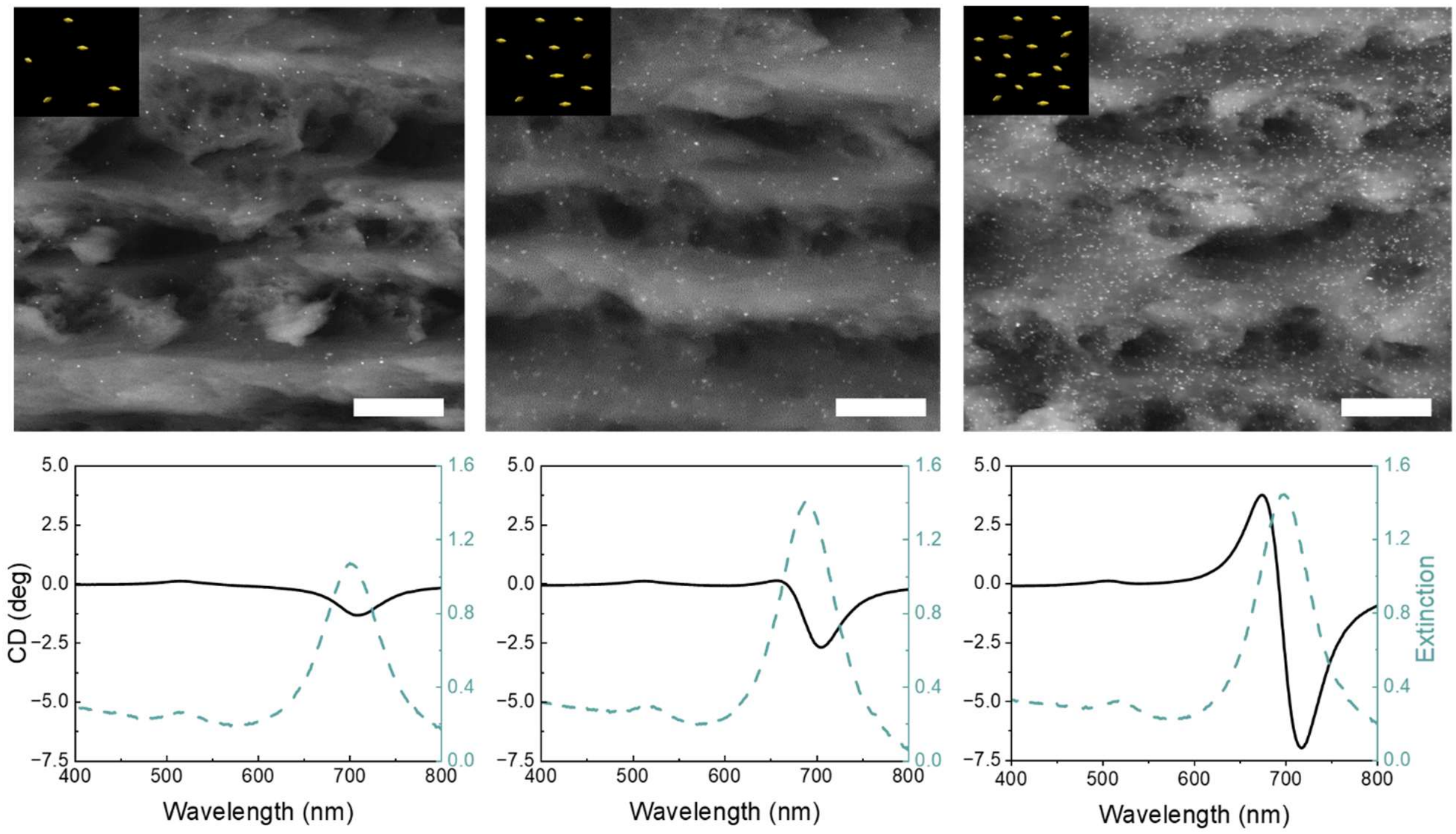

**Figure S5.** *Top*: SEM with backscattered electron detector of cross-section of films with different gold nanoparticle loadings (left to right [$Au^0$] ≈ 0.45, 0.9, 1.8 mg/mL) (AuBPs aspect ratio 2.2). Scale bar 2 μm. *Bottom:* Circular dichroism and extinction (dashed line) spectra showing the transition from a single to a bisignate signal of the dichroic bands centered around 700 nm when increasing the gold nanoparticle loading.

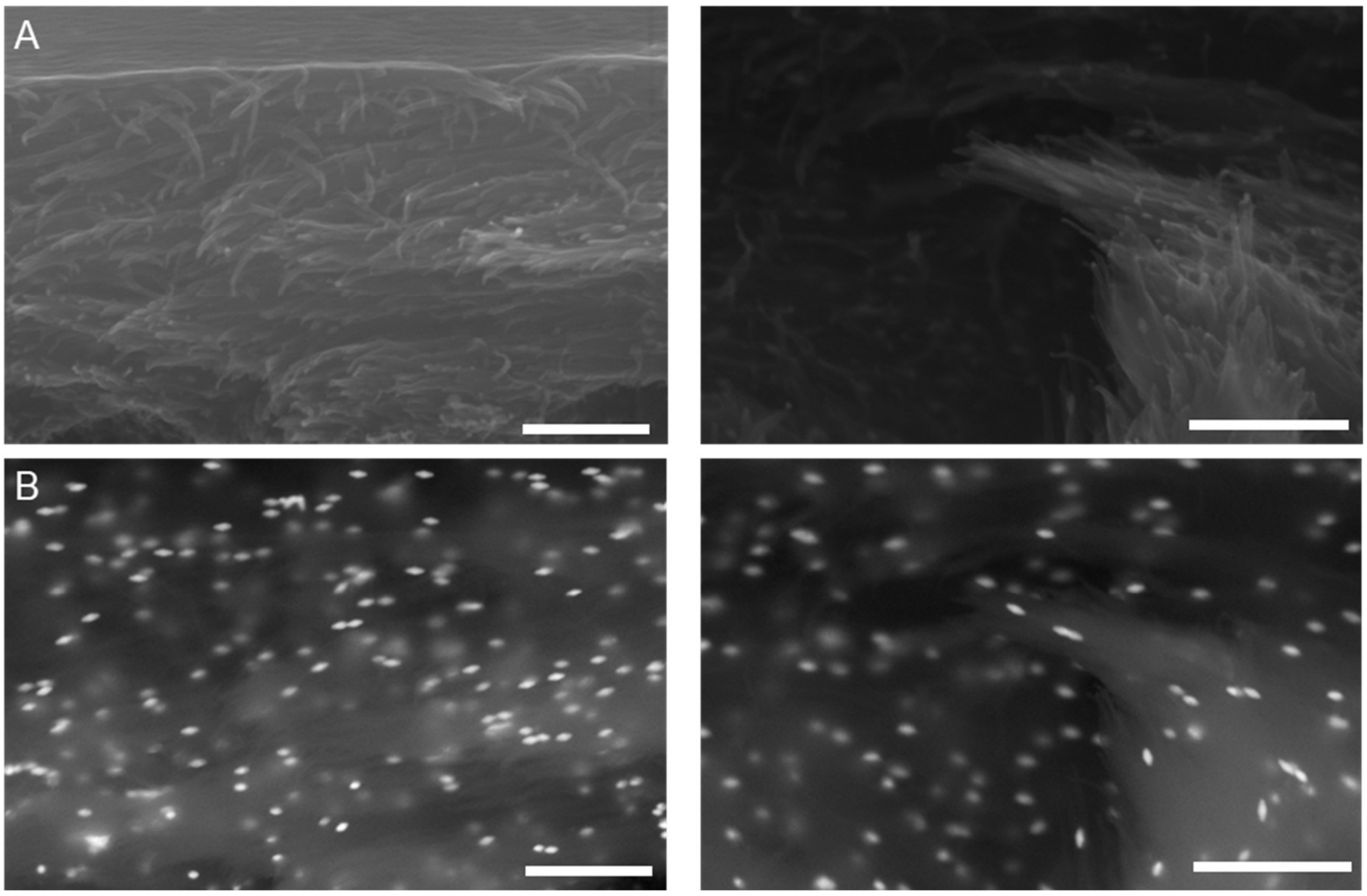

**Figure S6.** High-magnification SEM cross-section of a chitin helicoidal film containing gold nanoparticles (aspect ratio 2.2). (**A**) The figures present two different spots in the film, observed using a secondary electron detector that allows resolution of individual chitin nanocrystals. (**B**) The same two spots, when observed with a backscattered electron detector, reveal elongated nanoparticles within the film. Scale bars are 500 nm.

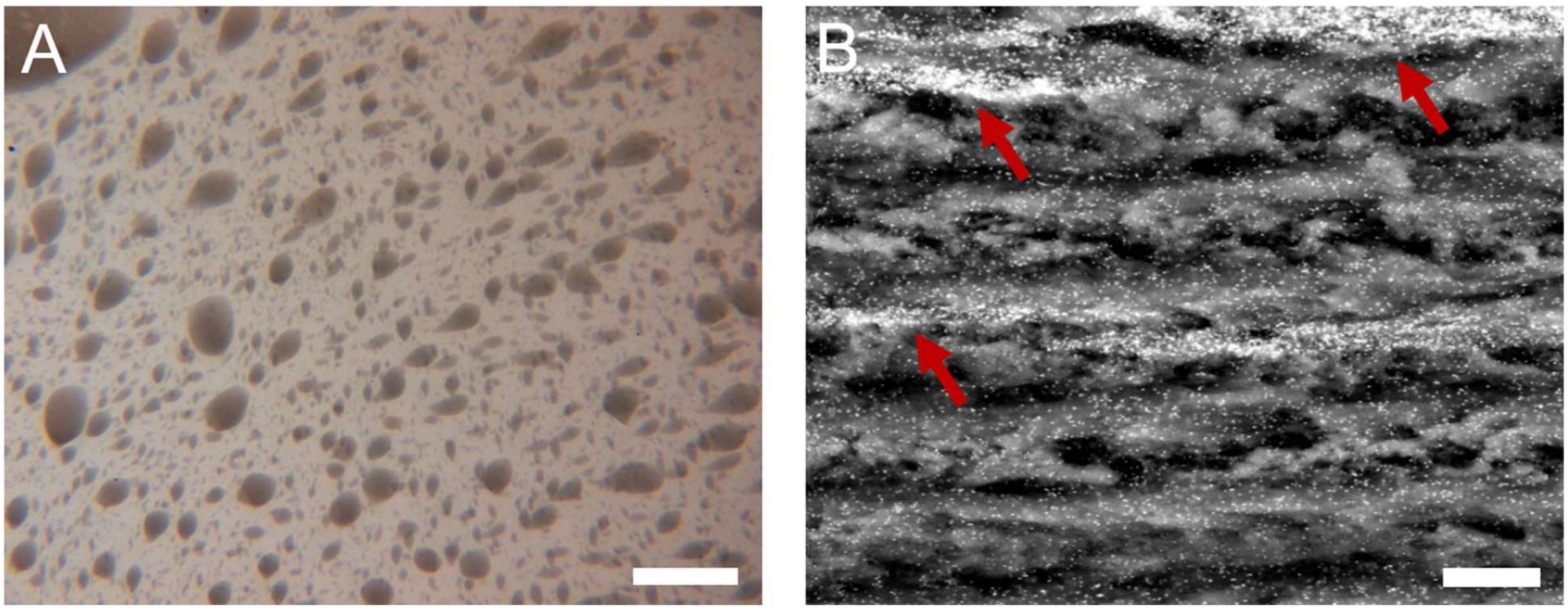

**Figure S7.** (**A**) Bright-field transmission microscopy image of a dry film (top view) showing the presence of a dark phase with a high concentration of gold nanoparticles (aspect ratio 3.2). Scale bar 100 μm. (**B**) Cross-sectional SEM image (backscattered electron detector) of a film, showing a high concentration of gold nanoparticles within this phase (highlighted by the red arrows). Vertical compression during drying causes these domains to appear flattened in the final film. Scale bar 5 μm.

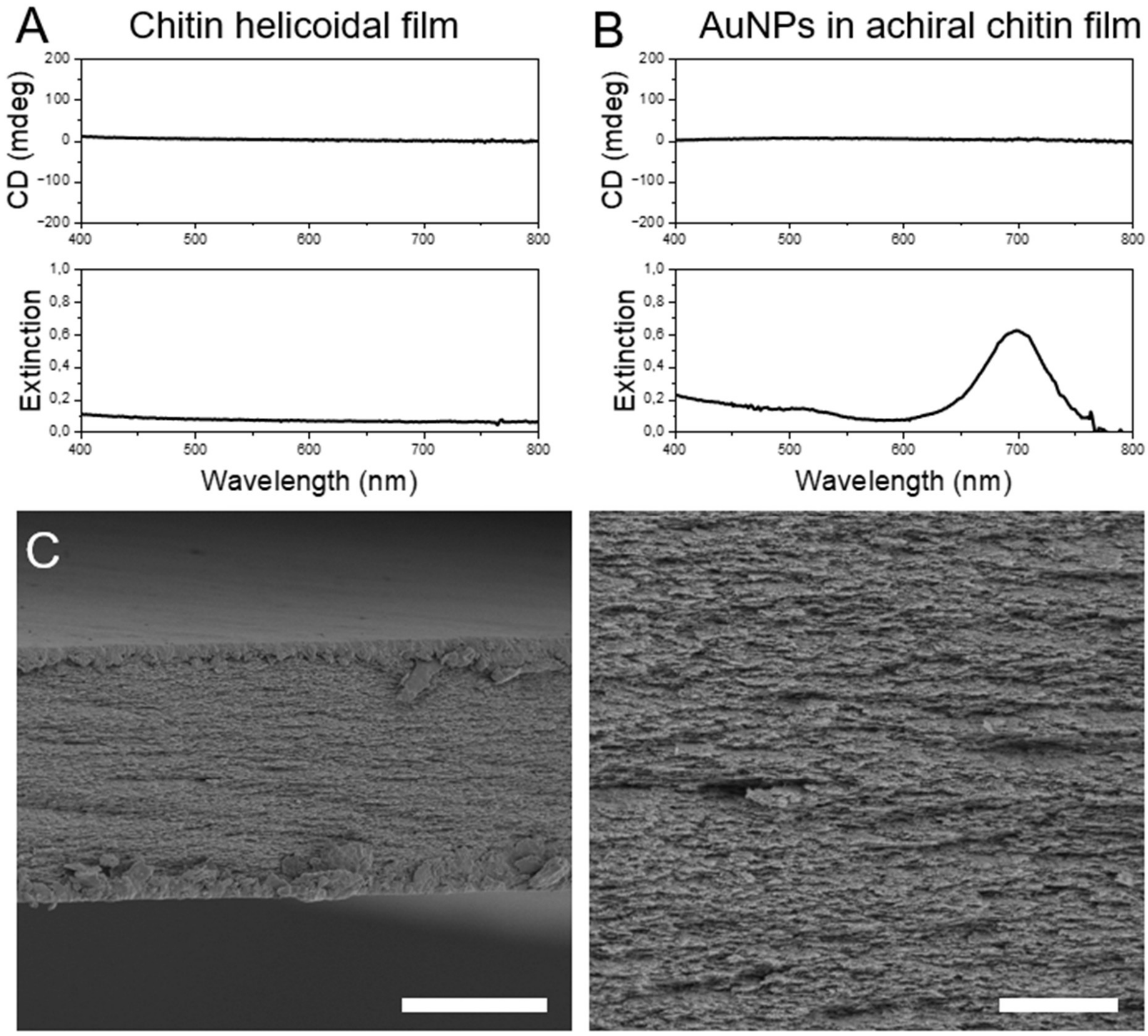

**Figure S8.** Circular dichroism and extinction spectra of control samples: (**A**) pure chitin helicoidal film and (**B**) achiral chitin film containing gold nanoparticles. In both cases, no circular dichroism is detected. (**C**) SEM cross-section at different magnifications of an achiral chitin film containing gold nanoparticles made by the isotropic phase of a ChNCs biphasic suspension, showing the absence of the helicoidal structure. Scale bars are 25 µm and 5 µm for the left and right images, respectively.

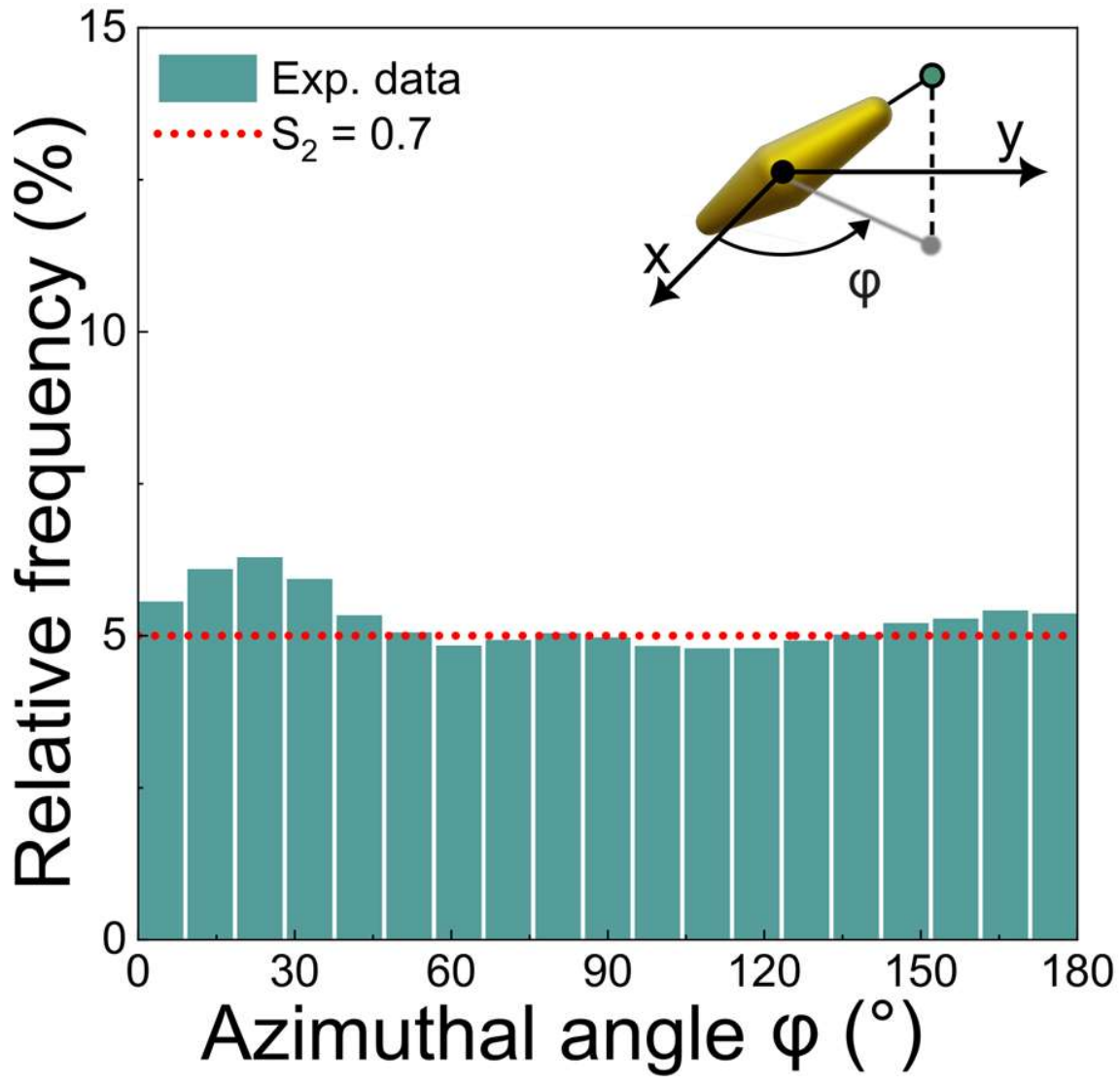

**Figure S9.** Azimuthal angle ($\varphi$) distribution of the tomographic dataset, along with the angle distribution expected for an $S_2$ parameter of 0.7. The azimuthal angle $\varphi$ distribution is uniform, as expected from a helicoidal structure where all the “rotations” are present.

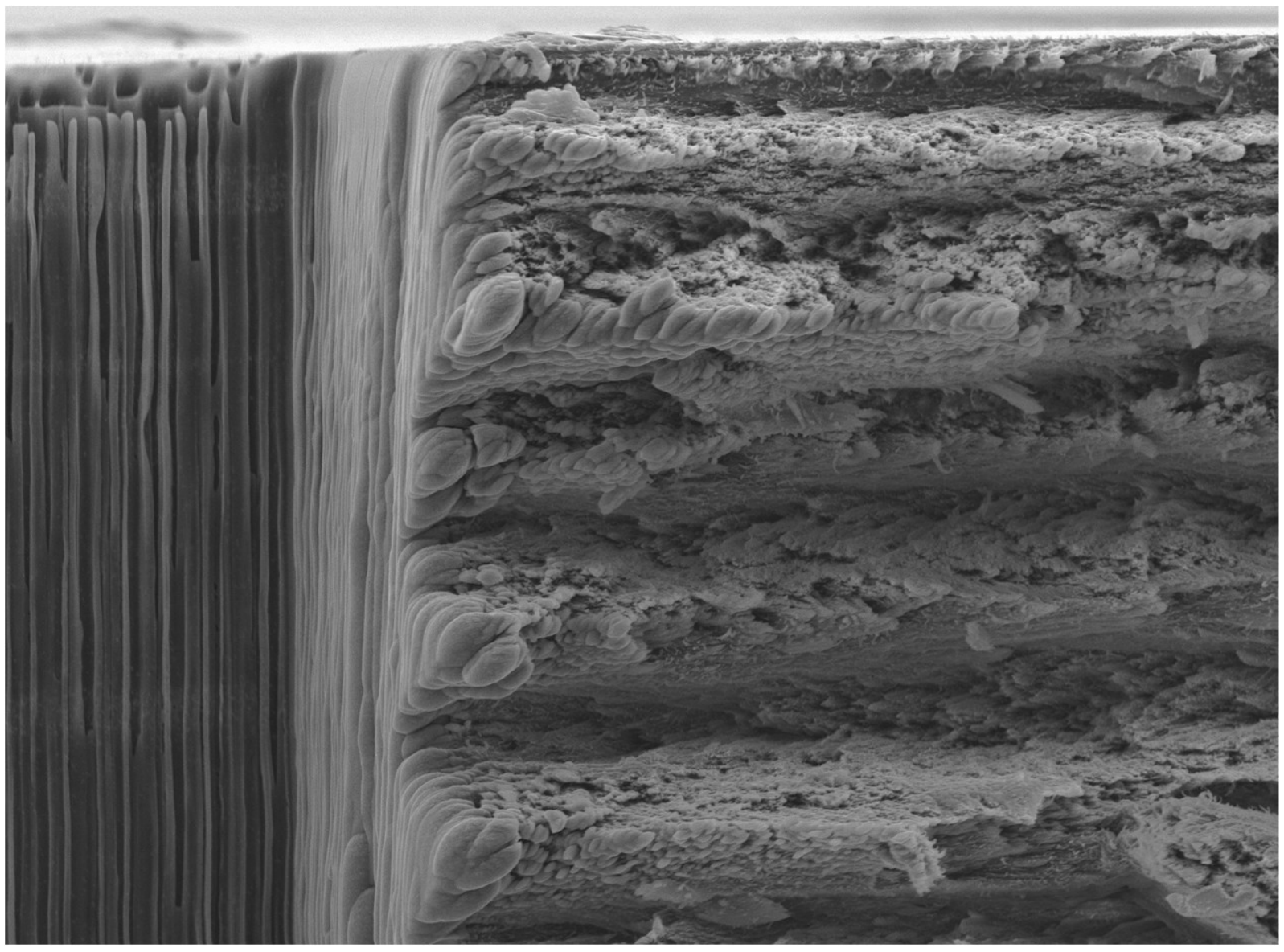

**Figure S10.** Scanning electron microscopy images of the cross-section recorded near the region (left side of the image) milled with FIB-SEM. Scale bar, 10 µm.

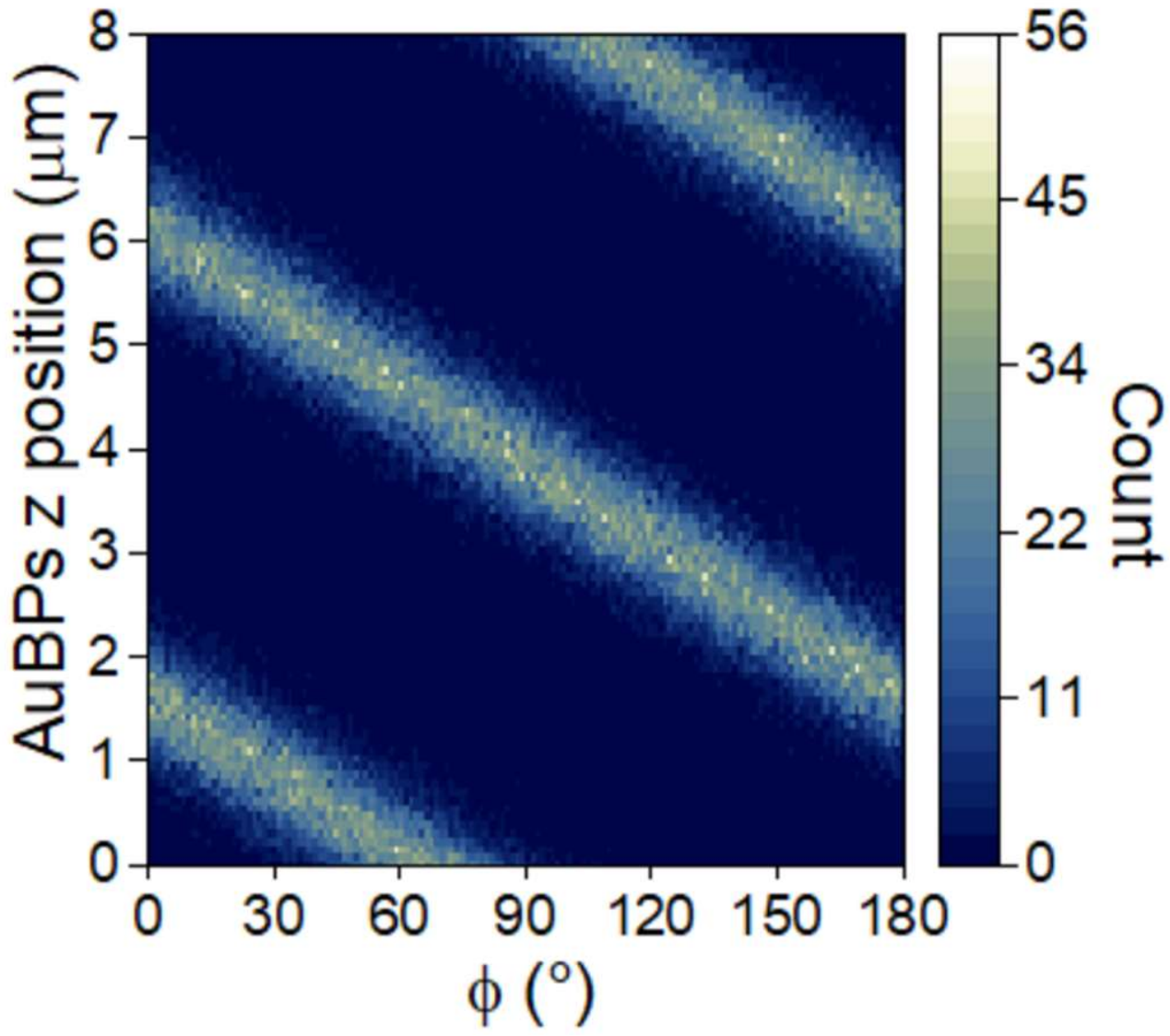

**Figure S11.** Simulated *φ-z* correlation diagram of ≈ 1.3 × $10^5$ particles with nematic order parameter $S_2 = 0.7$, and pitch of 9 μm generated using the *Helicoidal Model*

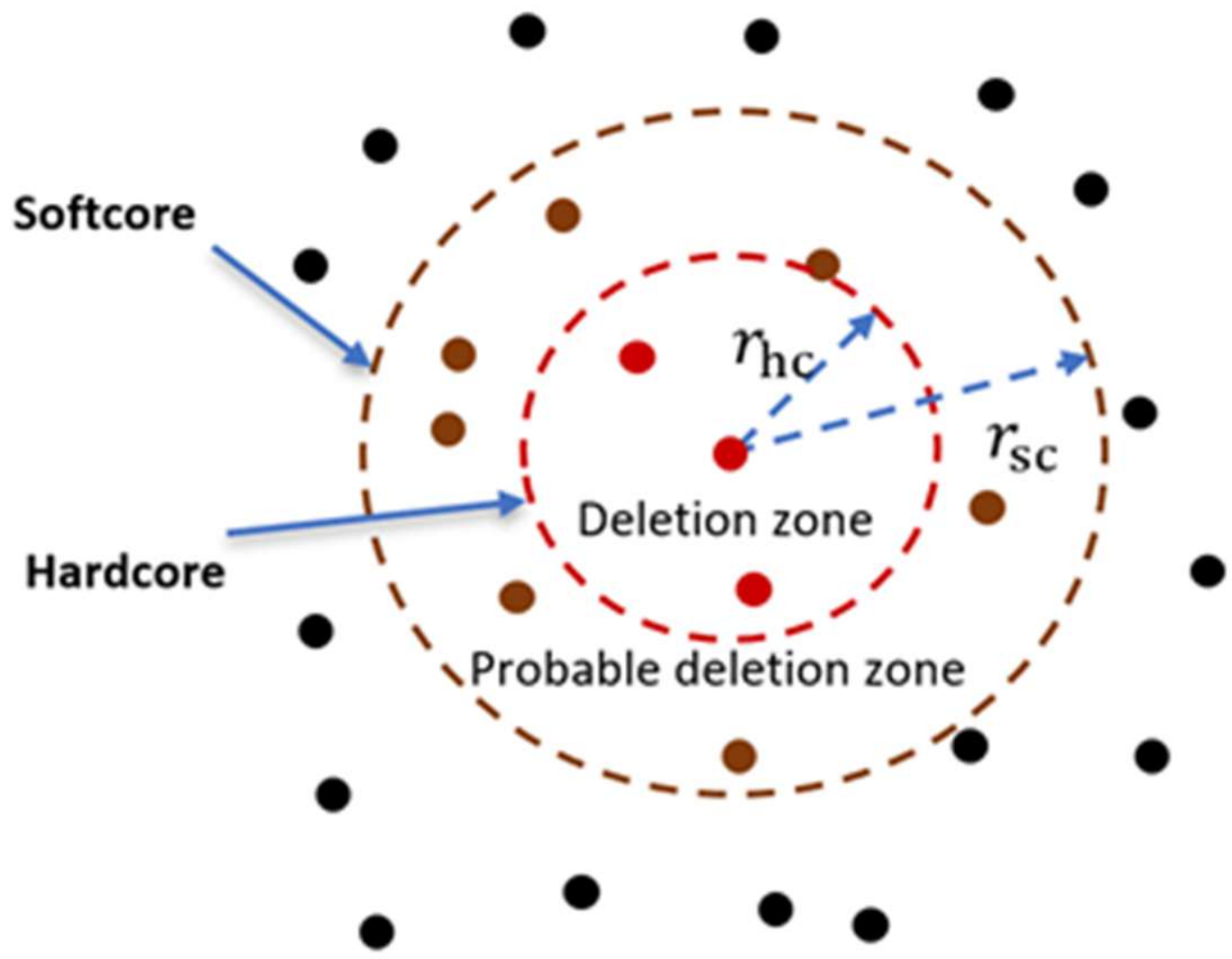

Deletion probability

$$f(r; r_{hc}, r_{sc}) = \begin{cases} 1 & 0 \leq r < r_{hc} \\ \dfrac{r_{sc} - r}{r_{sc} - r_{hc}} & r_{hc} \leq r < r_{sc} \\ 0 & r_{sc} \leq r \end{cases}$$

**Figure S12. Tailored Disorder Point Process.** Starting from a Poisson point distribution at the target density, points are selectively removed using a distance-dependent rule (*52*): (i) neighbors within the hard-core radius are always removed with probability one; (ii) neighbors between the hard-core and soft-core radii are removed with a probability that decreases linearly from one to zero as the distance approaches the soft-core limit. After the thinning process, new points are iteratively reinserted, up to a maximum number of attempts, to restore the target density.

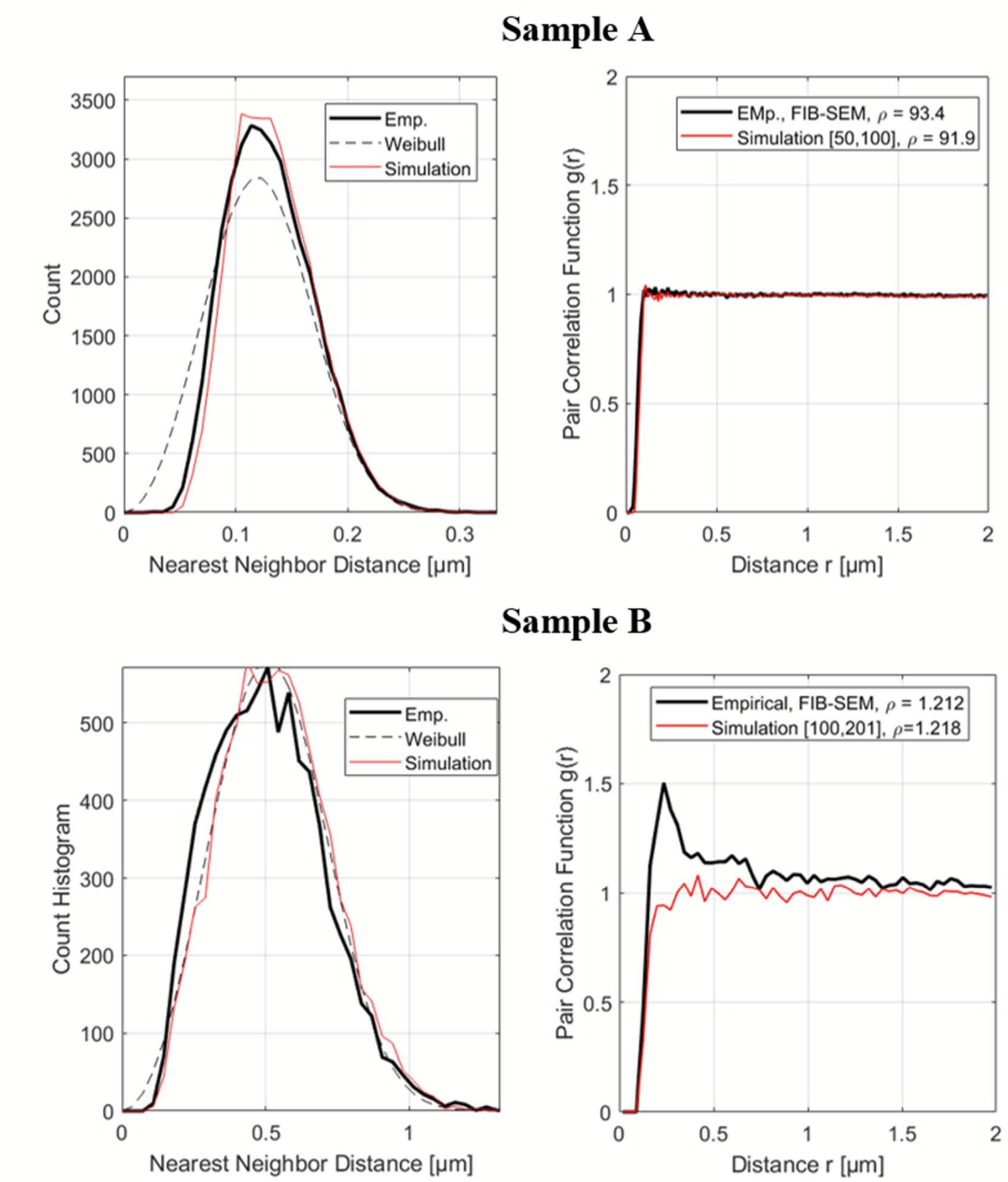

**Figure S13. Modeled point process.** Nearest-neighbor distributions and pair-correlation functions for the *Tomographic Dataset* and the modeled tailored disordered point process at two densities (upper and lower rows). The Weibull distribution denotes the nearest-neighbor distance for a Poisson point process and is shown for reference. Hard-core and soft-core interaction parameters were chosen to reproduce the structural statistics of two tomographic datasets, with parameters $r_{hc} = 35 \times \left(1 + 2\rho^{-1/3}\right)[nm]$, $r_{sc} = 70 \times \left(1 + 2\rho^{-1/3}\right)\,[nm]$.

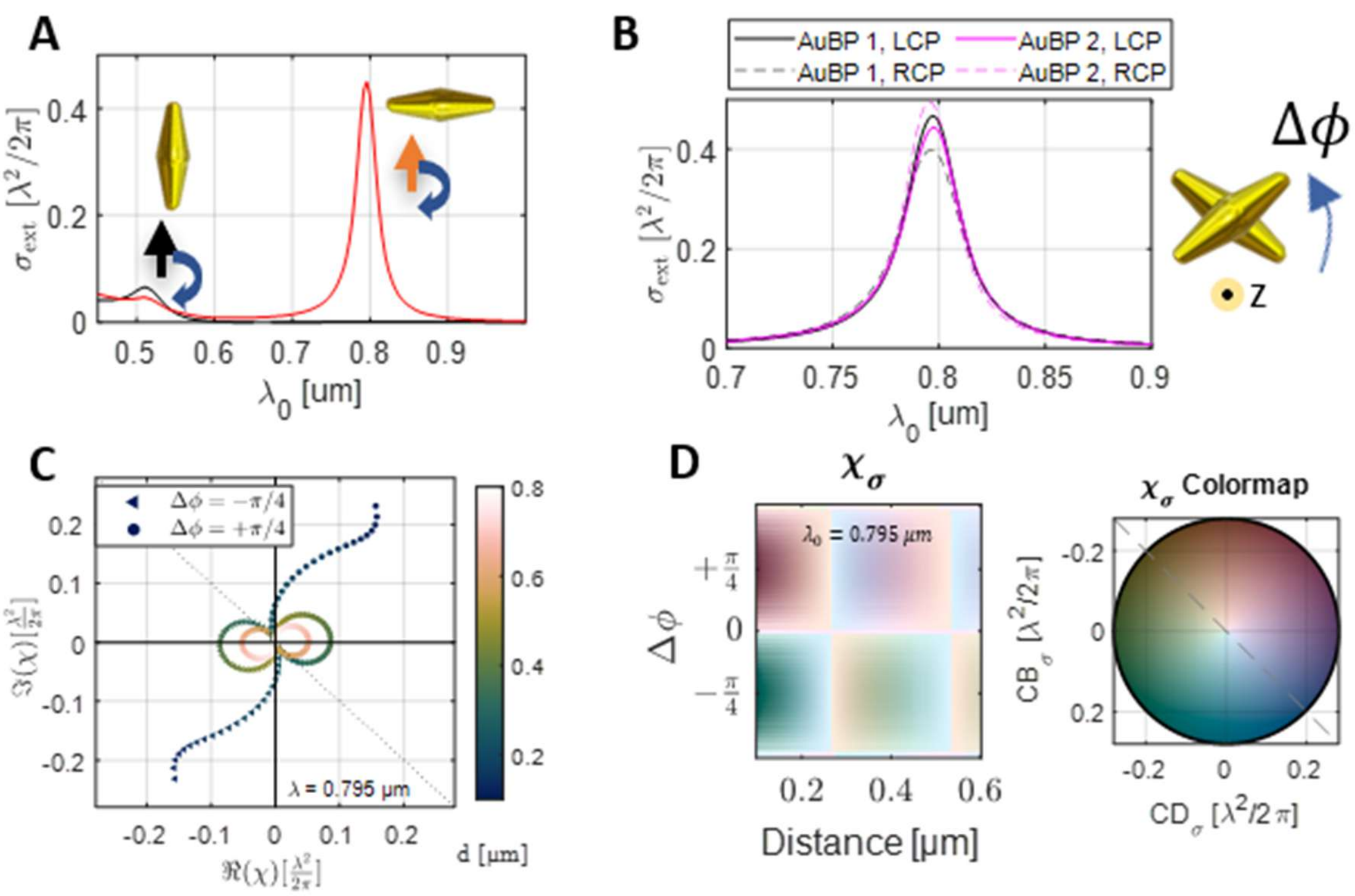

**Figure S14**. **Optical response of the single and dimer bipyramid**. (**A**) Extinction cross section of a single AuBP under left-circularly polarized illumination. Owing to the rotational symmetry of the particle, both polarizations produce identical optical responses. The two resonances correspond to the longitudinal and transverse modes. (**B**) Extinction cross sections of two coupled AuBPs forming a dimer with interparticle distance (*d*) and relative azimuthal orientation under circularly polarized illumination propagating along the (*z*)-axis. The individual extinction contributions of each particle are shown separately. (**C**) Complex-plane representation of the chiroptical index ($\chi_\sigma$) at resonance ($\lambda_0 = 795$ nm) for enantiomeric dimers as a function of interparticle distance. (**D**) Chirality-phase map of the dimer at $\lambda_0 = 795$ nm as a function of distance and relative orientation. The bivariate colormap encodes the full complex chiroptical response, with hue representing the chirality phase and color saturation the chiroptical strength.

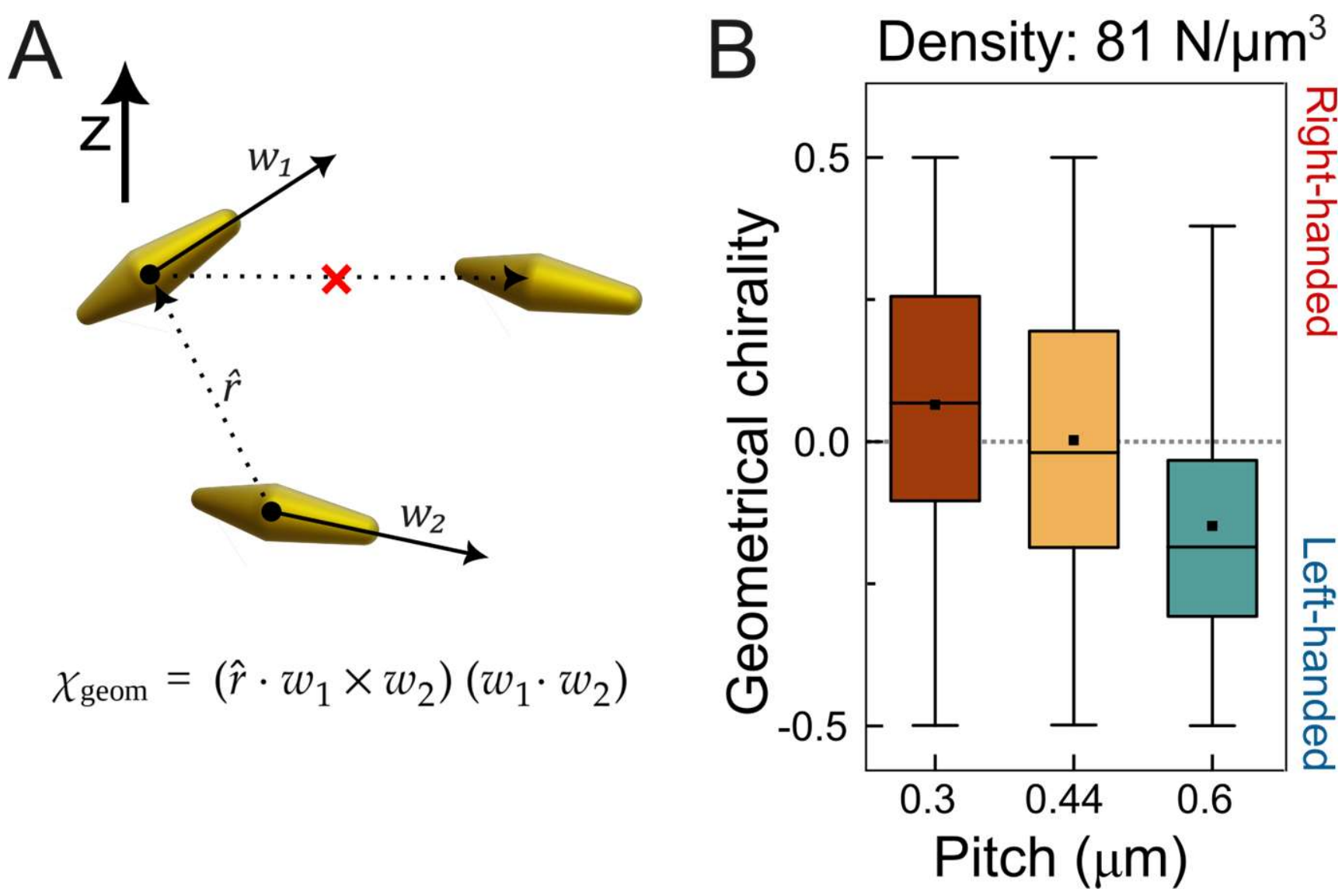

**Figure S15. Geometrical chirality.** (**A**) Schematic of the geometrical chirality index $\chi_{geom}$ calculation. Particles on the same *z*-plane are excluded from the calculation as they do not contribute to the chiroptical response. (**B**) Geometrical chirality distribution calculated for $1.3 \times 10^5$ particles generated with the *Helicoidal model* with a density of 81 N/µm$^3$ and different pitches. The boxplot shows the mean (square), median (line), Q1-Q3 (box), and 1.5 IQR (whiskers).

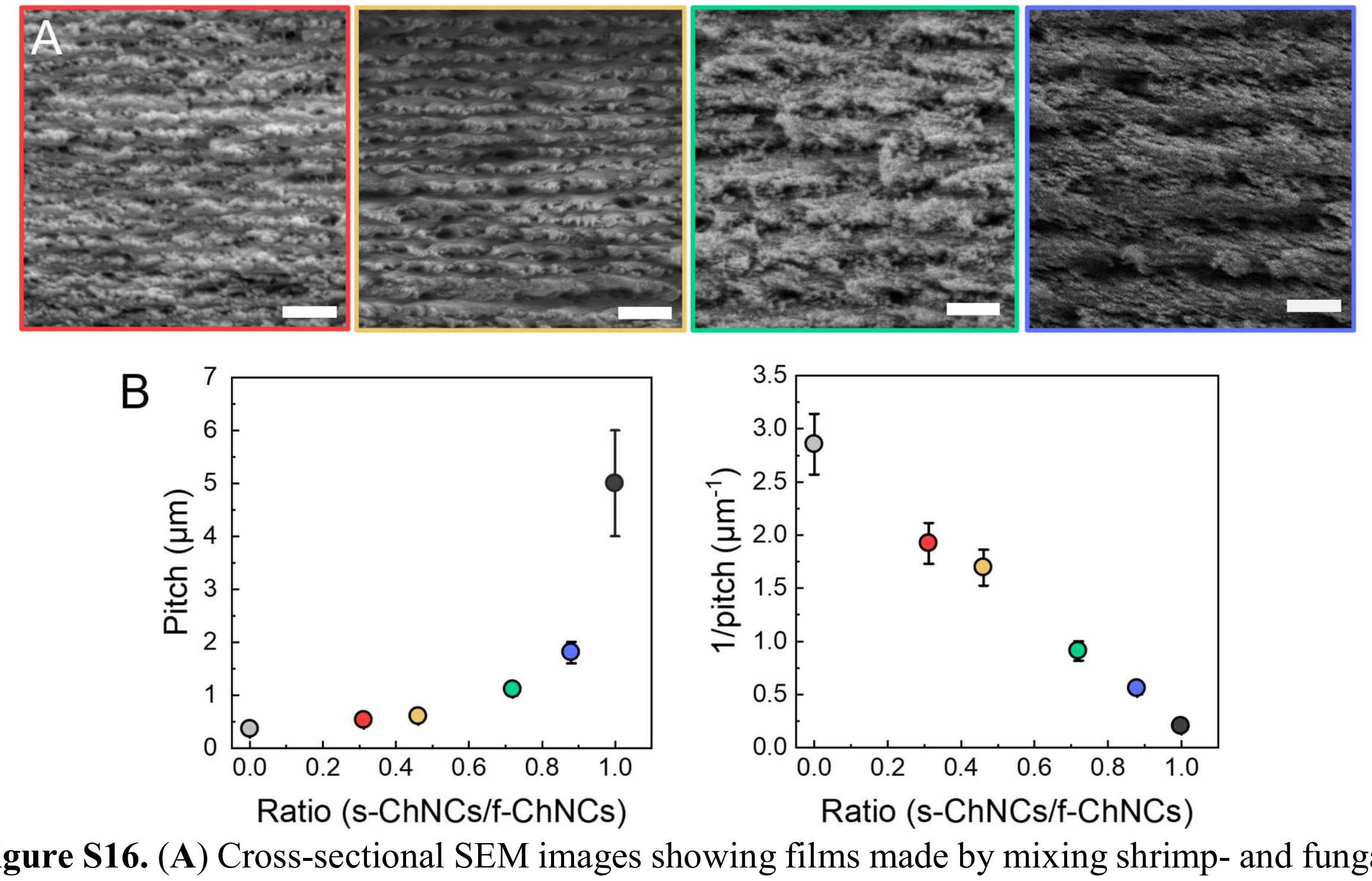

**Figure S16.** (**A**) Cross-sectional SEM images showing films made by mixing shrimp- and fungal-derived chitin nanocrystals in different weight ratios (left to right: 0.31, 0.46, 0.72, 0.88). Scale bars 1 µm. (**B**) Plots showing the measured variation of pitch in the solid films at different ratios of shrimp- and fungal-derived chitin nanocrystals. Error bars are the associated standard deviation.

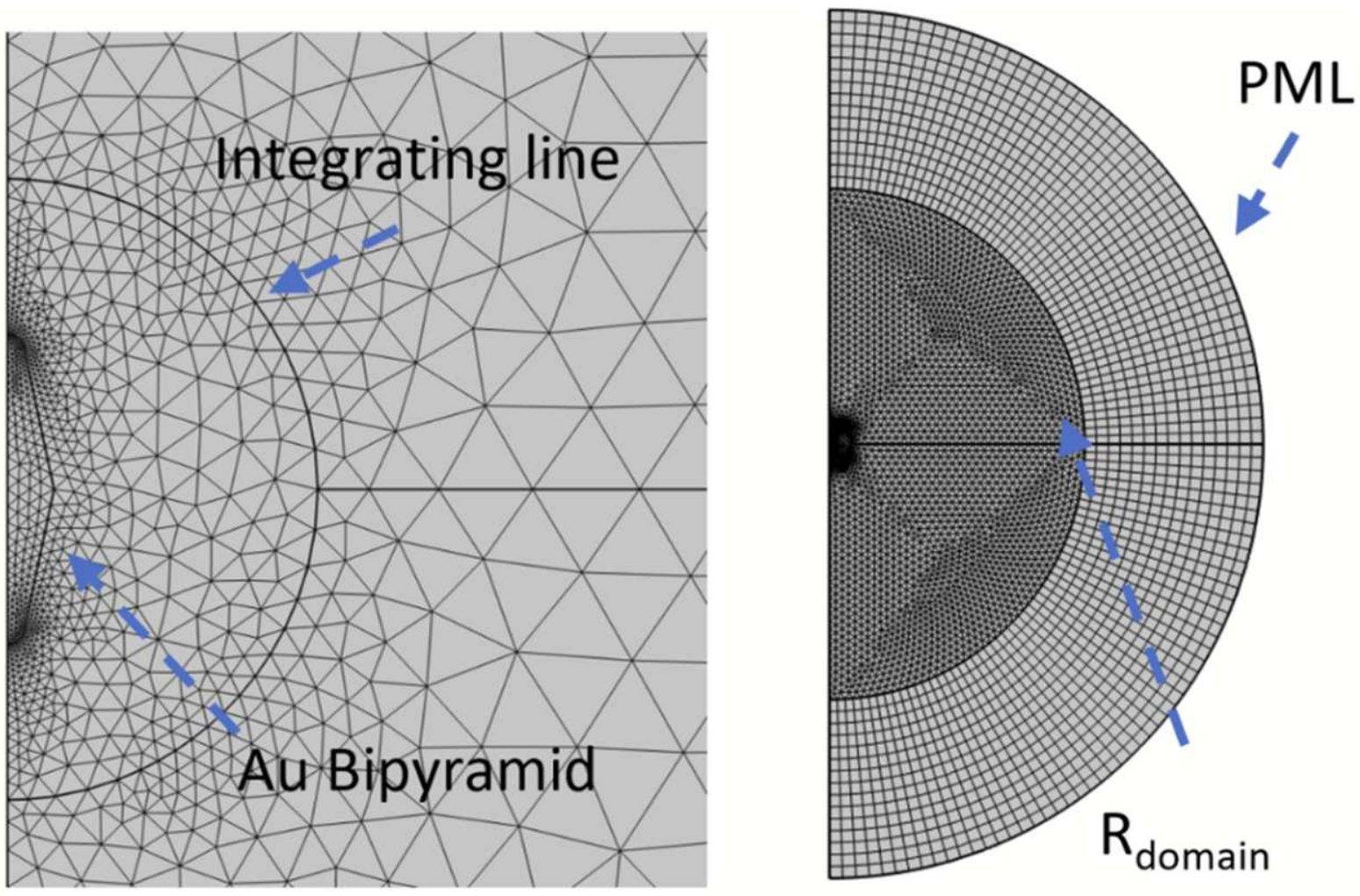

**Figure S17**. **T-matrix retrieval of a single AuBP using COMSOL Multiphysics**. The computational mesh used to derive the T-matrix of an individual gold bipyramid (AuBP). The dashed integration contour indicates the surface over which the scattered field is projected onto vector spherical wave functions following multipole excitation (see T-matrix calculation of an individual bipyramid). The AuBP is approximated as a biconical shape with a spherical tip. The rotational symmetry is exploited to substantially reduce the computational cost of the T-matrix retrieval. The particle geometry, including its dimensions and tip curvature, was adjusted to reproduce the experimentally observed plasmonic resonances, consistent with previously reported morphologies of gold bipyramids.

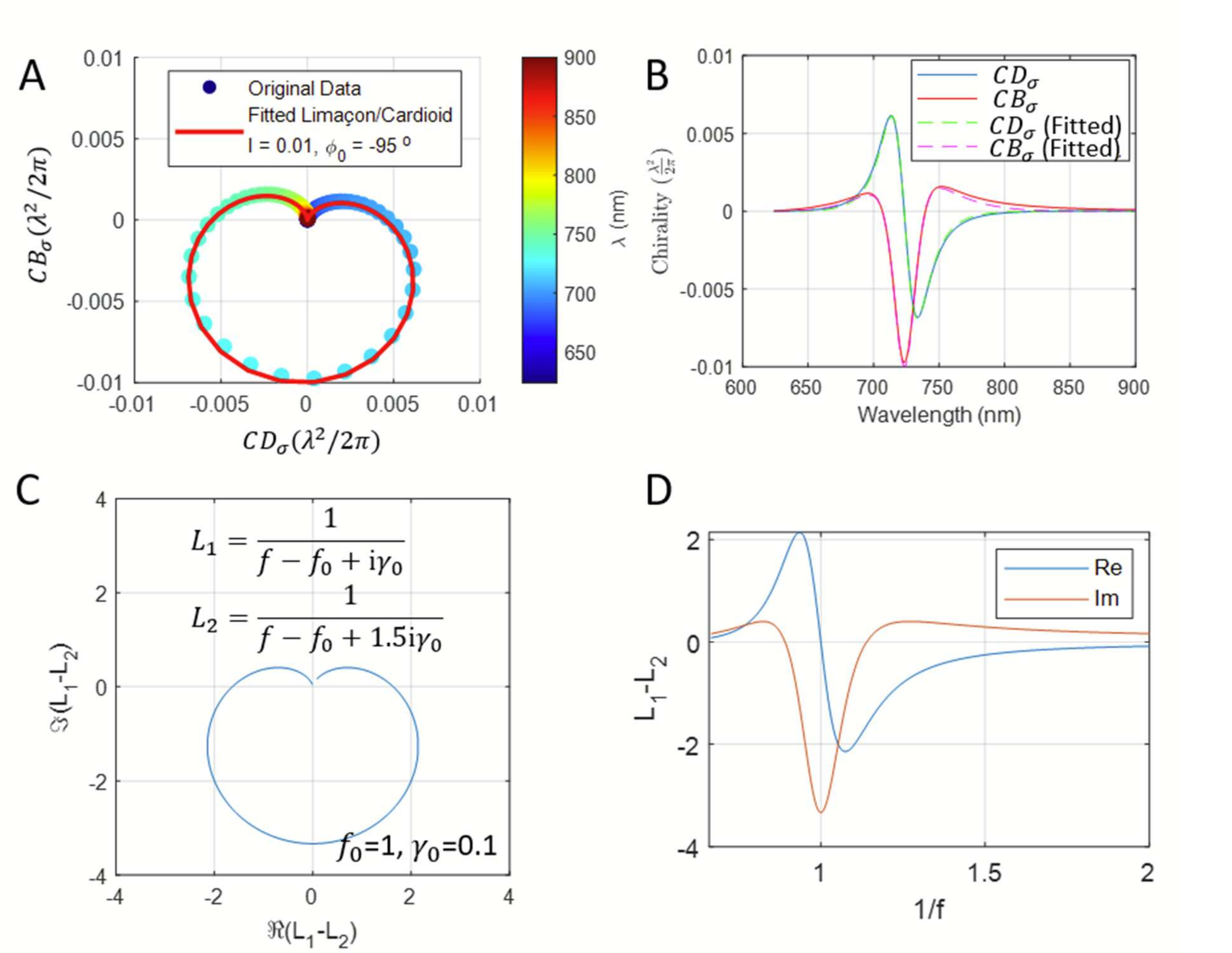

**Figure S18**. **Limaçon model and differential Lorentzian representation of the chiroptical response. (A)** Complex-plane representation of the chiroptical index $\chi_\sigma$ obtained from the simulated tomographic dataset together with a fitted limaçon model. The trajectory is well described by $r(t) = I[1 + \cos(t - \phi_0)]$, $\chi_\sigma(t) = r(t)e^{it}$. The color scale denotes the wavelength. The fitted trajectory corresponds to a cardioid, a special case of the limaçon family. **(B)** Comparison between the simulated $CD_\sigma$ and $CB_\sigma$ spectra and those reconstructed from the fitted limaçon model, showing excellent agreement. **(C)** Illustration that a limaçon trajectory can be generated from the difference of two Lorentzian resonances with identical resonance frequencies but different damping factors. **(D)** Real and imaginary parts of the differential Lorentzian response, L1-L2. The resulting spectral line shapes closely resemble the circular dichroism and circular birefringence spectra obtained from the simulated *Tomographic Dataset*.

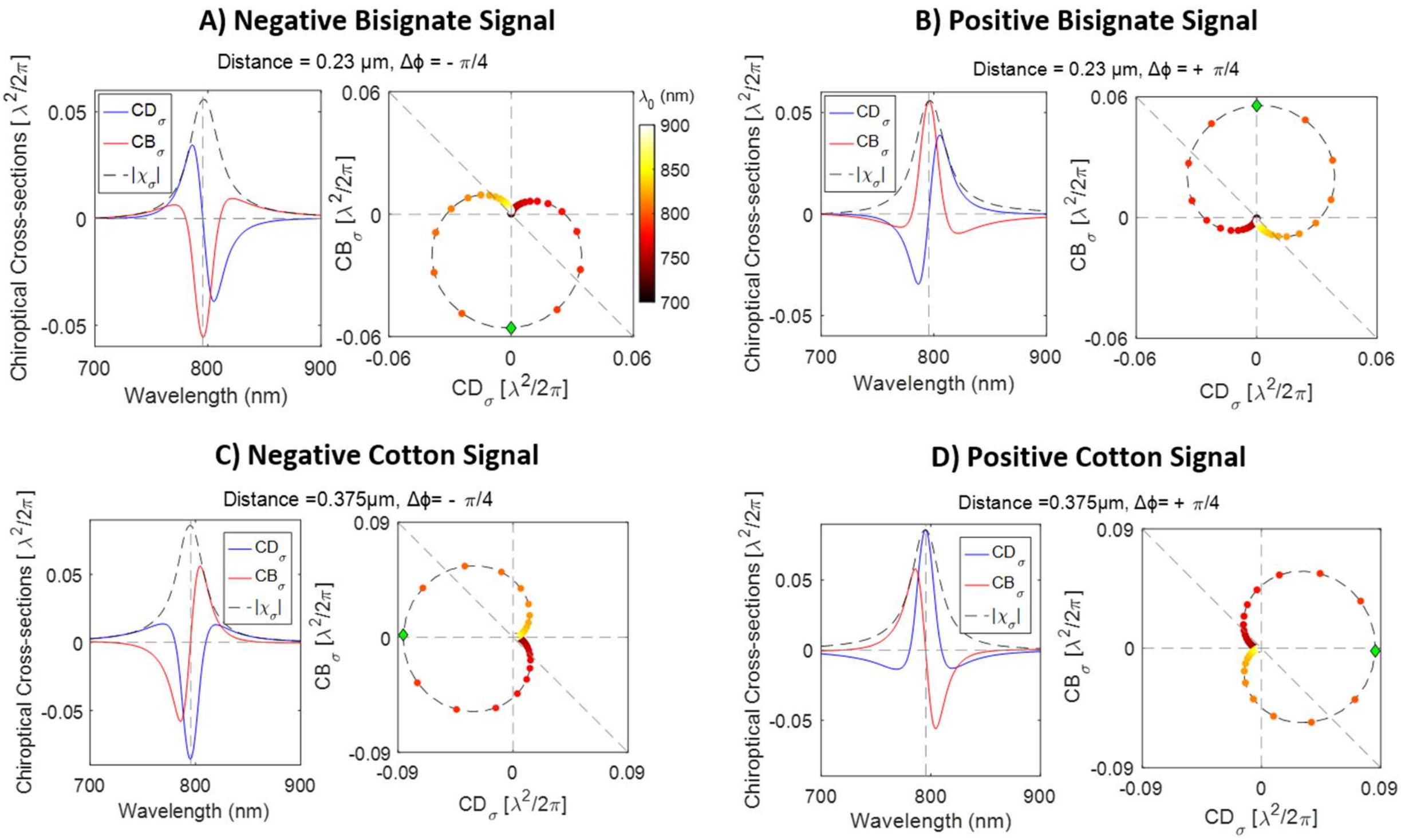

**Figure S19**. **Spectral shape and chiroptical index representation in the complex plane.** Chiroptical index ($\chi_\sigma$) simulation of a dimer shown in fig. S14 for different interparticle distances and relative orientations using the T-matrix approach. The chiroptical index ($\chi_\sigma = CD_\sigma + \mathrm{i}\, CB_\sigma$ ) forms a limaçon trajectory in the complex plane. Green diamonds indicate the spectral position of the plasmonic resonance. The color bar shows the clockwise wavelength progression. The left panel shows the real (i.e., $CD_\sigma$) and the imaginary (i.e., $CB_\sigma$) and magnitude ($|\chi_\sigma|$) spectra of the chiroptical index $\chi_\sigma$ while the right figure shows the complex representation. The spectral shape of a resonance can be predicted by tracing the $\chi_\sigma$ at resonance. The figures show (**A**) negative bisignate signal, **(B)** positive bisignate signal, (**C**) negative Cotton signal, and (**D**) positive Cotton signal, respectively.

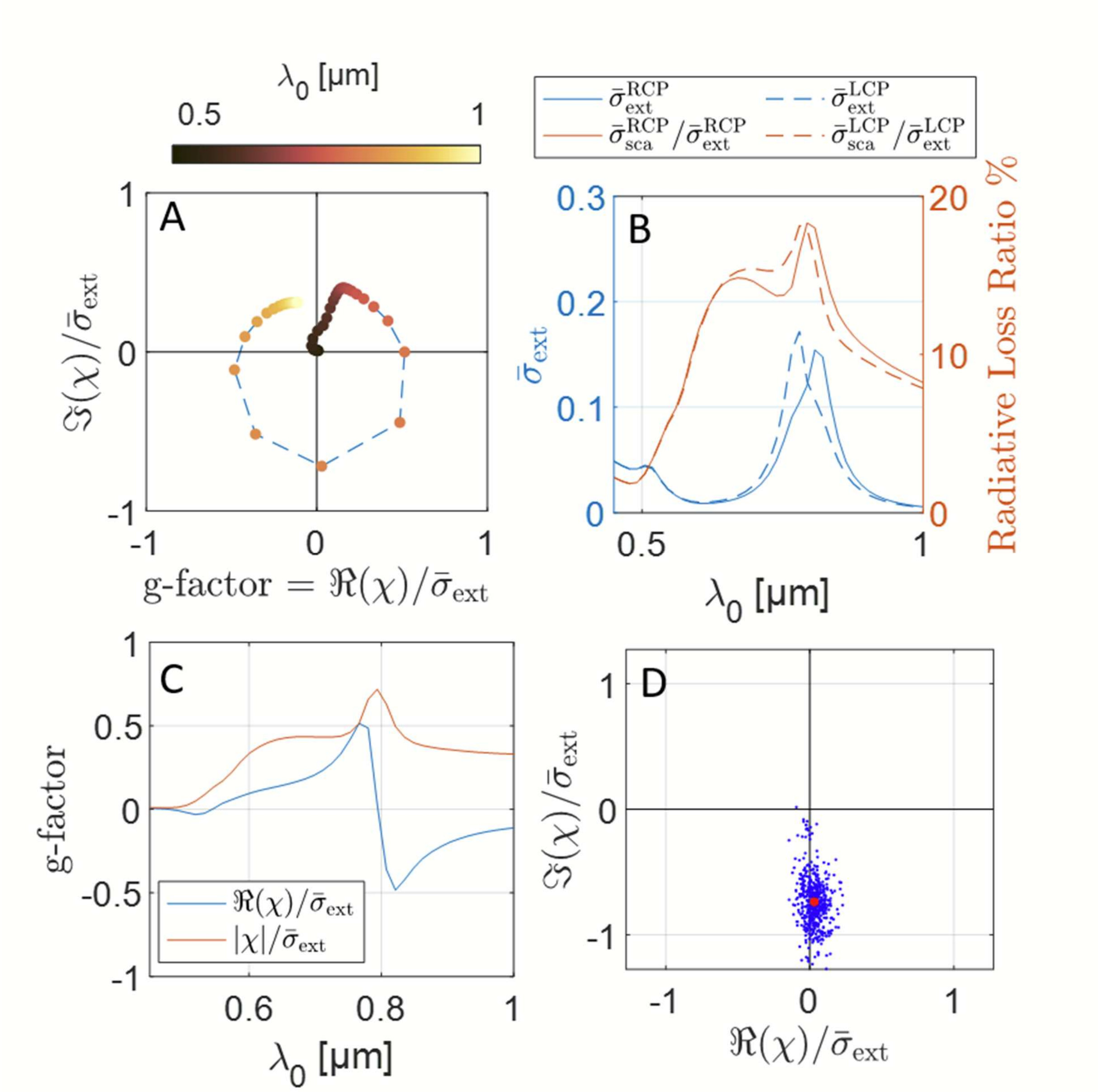

**Figure S20. Simulation of the *Tomographic Dataset*.** **(A)** Chiroptical index normalized by the polarization-averaged extinction cross section, shown in the complex plane; the real axis corresponds to the chiroptical dissymmetry g-factor. **(B)** Particle-averaged extinction cross section and radiative loss fraction of the system. **(C)** Spectra of the g-factor and the normalized chiroptical index, highlighting the limitation of g in capturing only the differential extinction component of the chiroptical response (CD), while $\chi_\sigma$ incorporates both CD and circular birefringence (CB). **(D)** Distribution of $\chi_\sigma$ across all simulation boxes. In total, 47,663 nanoparticles across 510 boxes were simulated, with coordinates and orientations taken from the *Tomographic Dataset*.

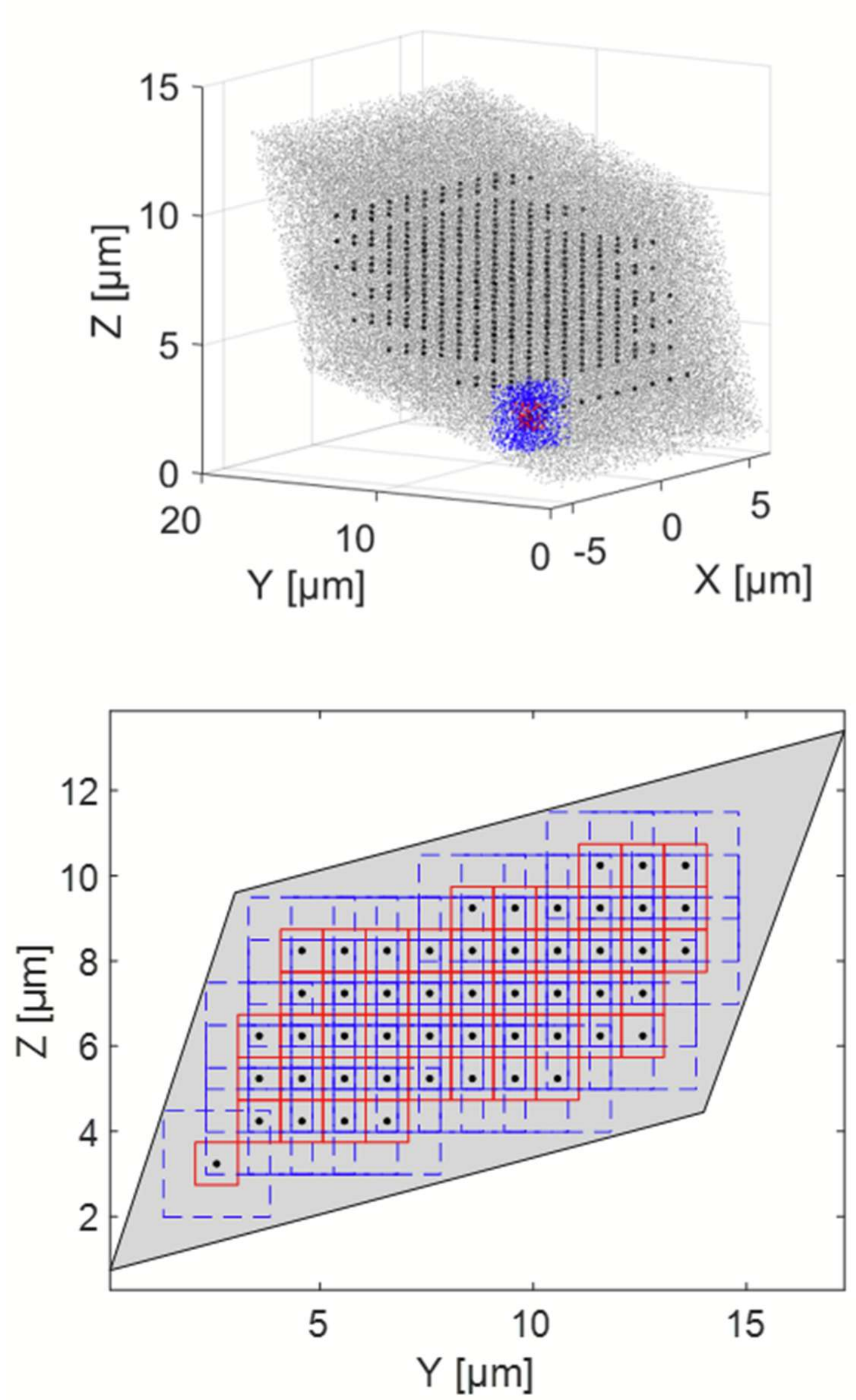

**Figure S21. Spatial subdivision of the *Tomographic Dataset* used for chiroptical simulations.** Gray dots represent AuBP coordinates obtained from FIB-SEM tomography and black dots the centers of the simulation boxes. The red volume indicates the region contributing to the chiroptical index ($\chi_\sigma$), while the surrounding blue volume denotes the buffer region used to mitigate boundary effects. In total, 510 boxes containing 47,663 AuBPs were analyzed, using a box size of 1 µm and a buffer thickness of 0.75 µm.

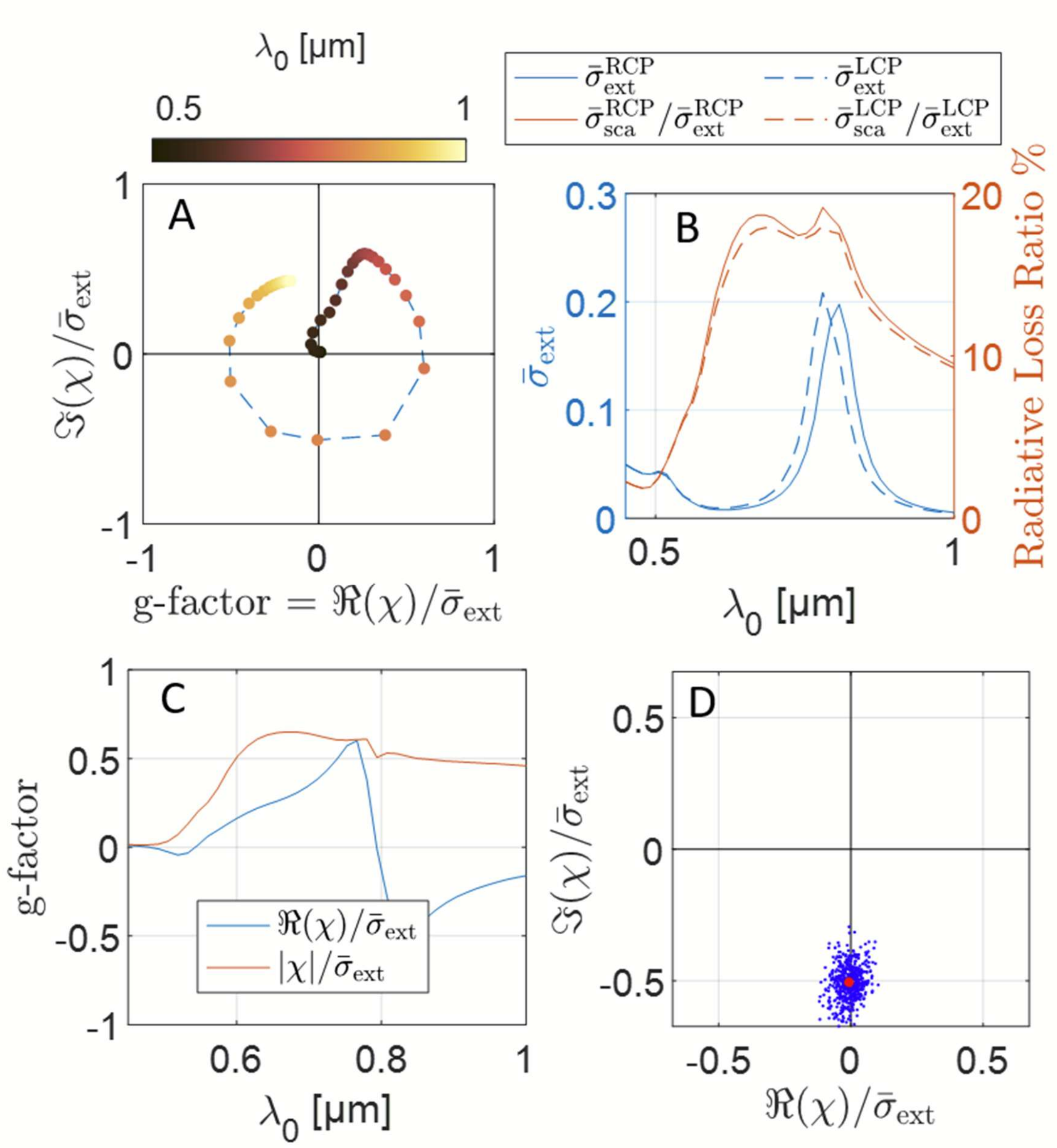

**Figure S22. Simulation of the *Tomographic Dataset* with orientation correction.** (**A**) Chiroptical index normalized by the polarization-averaged extinction cross section, shown in the complex plane; the real axis corresponds to the chiroptical dissymmetry g-factor. (**B**) Particle-averaged extinction cross section and radiative loss fraction of the system. (**C**) Spectra of the g-factor and the normalized $\chi_\sigma$. (**D**) Distribution of $\chi_\sigma$ across all simulation boxes. In total, 47,663 nanoparticles across 510 boxes were simulated, with coordinates taken from the *Tomographic Dataset*. Particle orientations were replaced by an ideal helicoidal configuration, with all AuBPs lying in the xy-plane and azimuthal angles following the helical pitch.

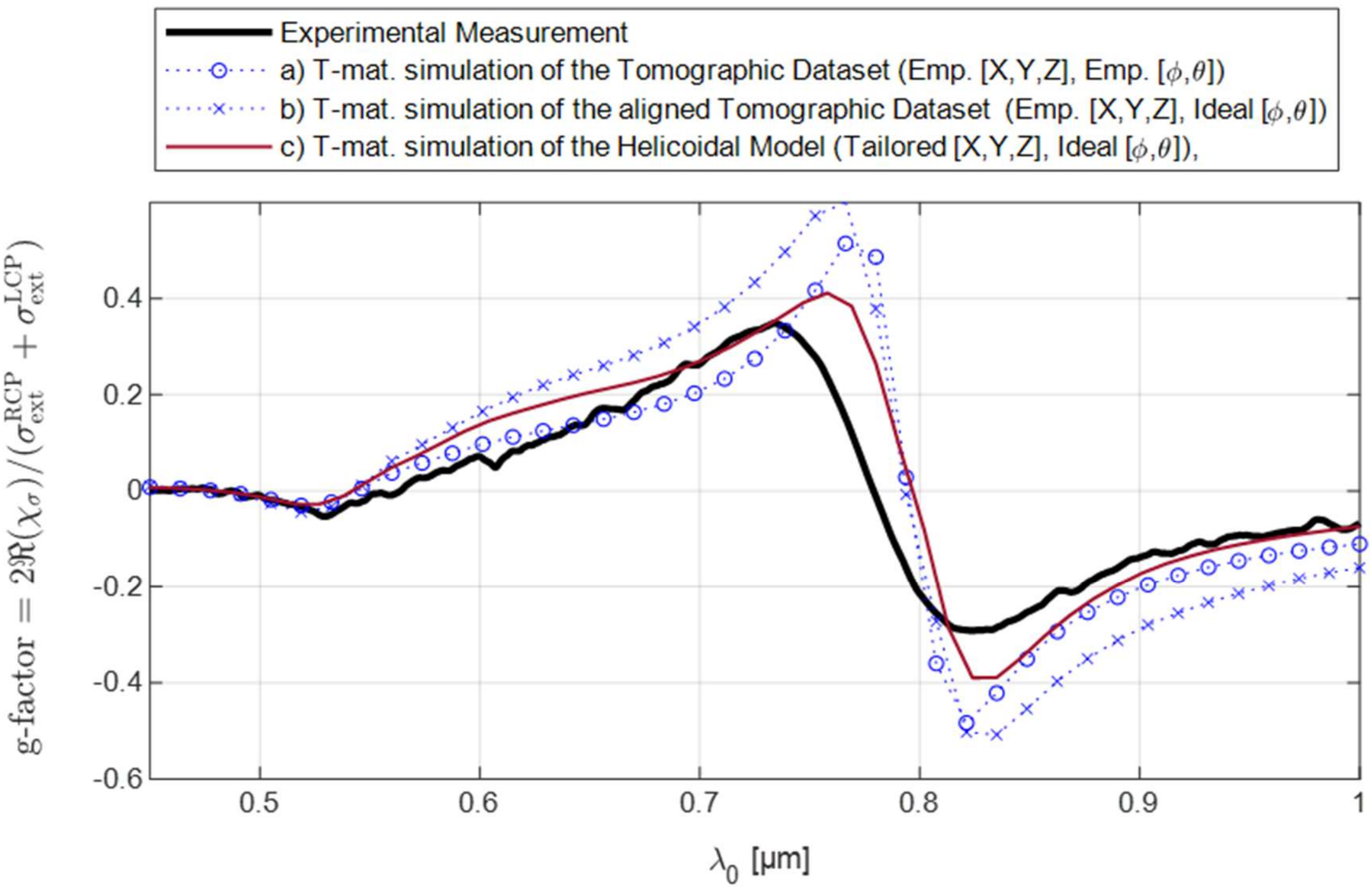

**Figure S23. Experiment and simulation of the *Tomographic Dataset* and models.** Comparison of experimentally measured g-factor spectra with three simulation approaches: (a) full *Tomographic Dataset input* (empirical coordinates and orientations), (b) *Tomographic Dataset* with orientations corrected to an ideal helicoidal configuration ($S_2 = 1$) and (c) *Helicoidal Model:* a tailored disordered point process for coordinates combined with ideal helicoidal orientations. All spectra show strong agreement, validating both the simulations and the underlying structural models.

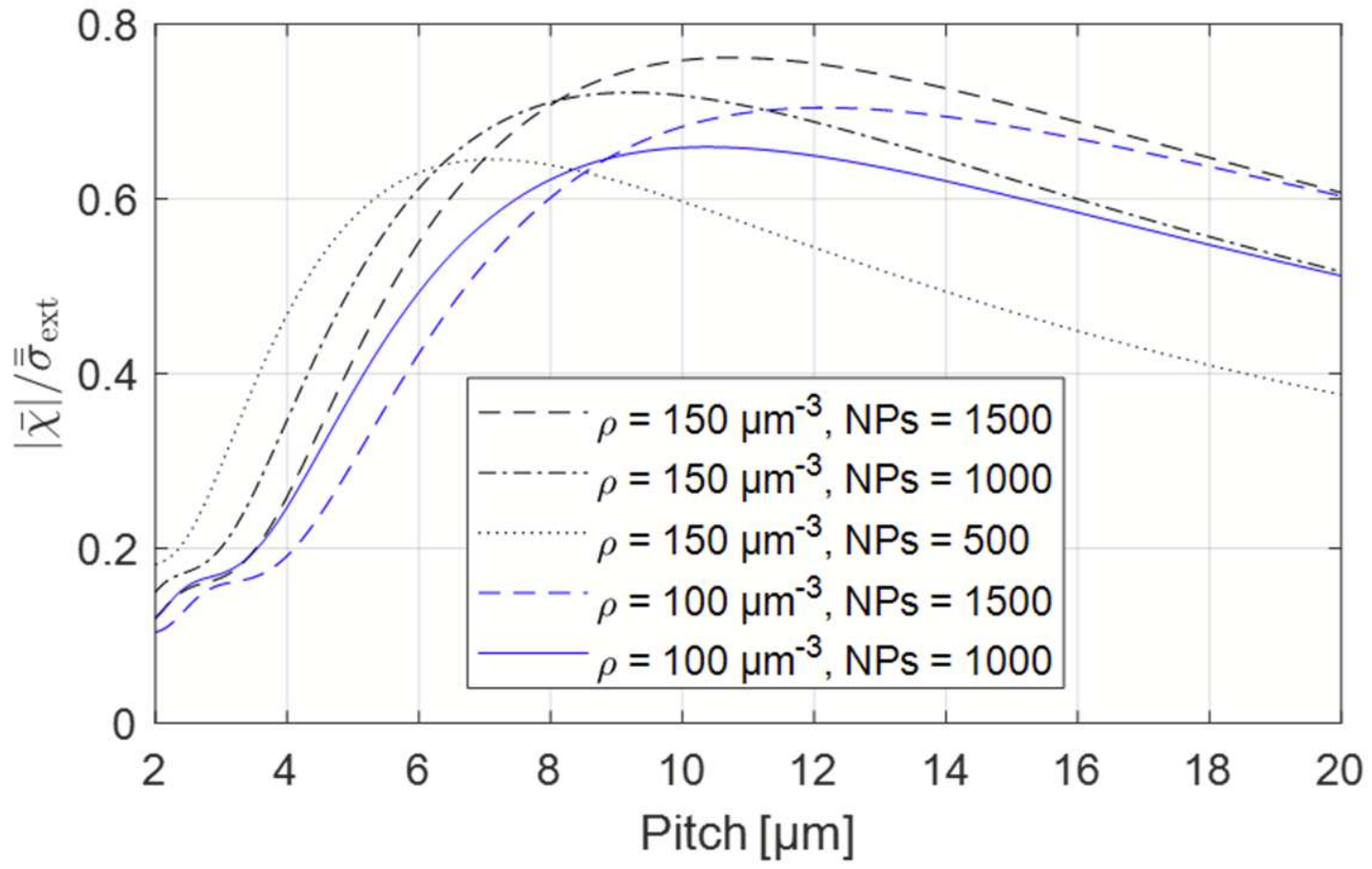

**Figure S24. Optimal pitch for maximizing the chiroptical response.** Normalized chiroptical index magnitude as a function of helicoidal pitch for various particle densities and ensemble sizes in the *Helicoidal Model*. A well-defined optimum pitch emerges for each configuration, corresponding to the maximum chiroptical efficiency of the system.

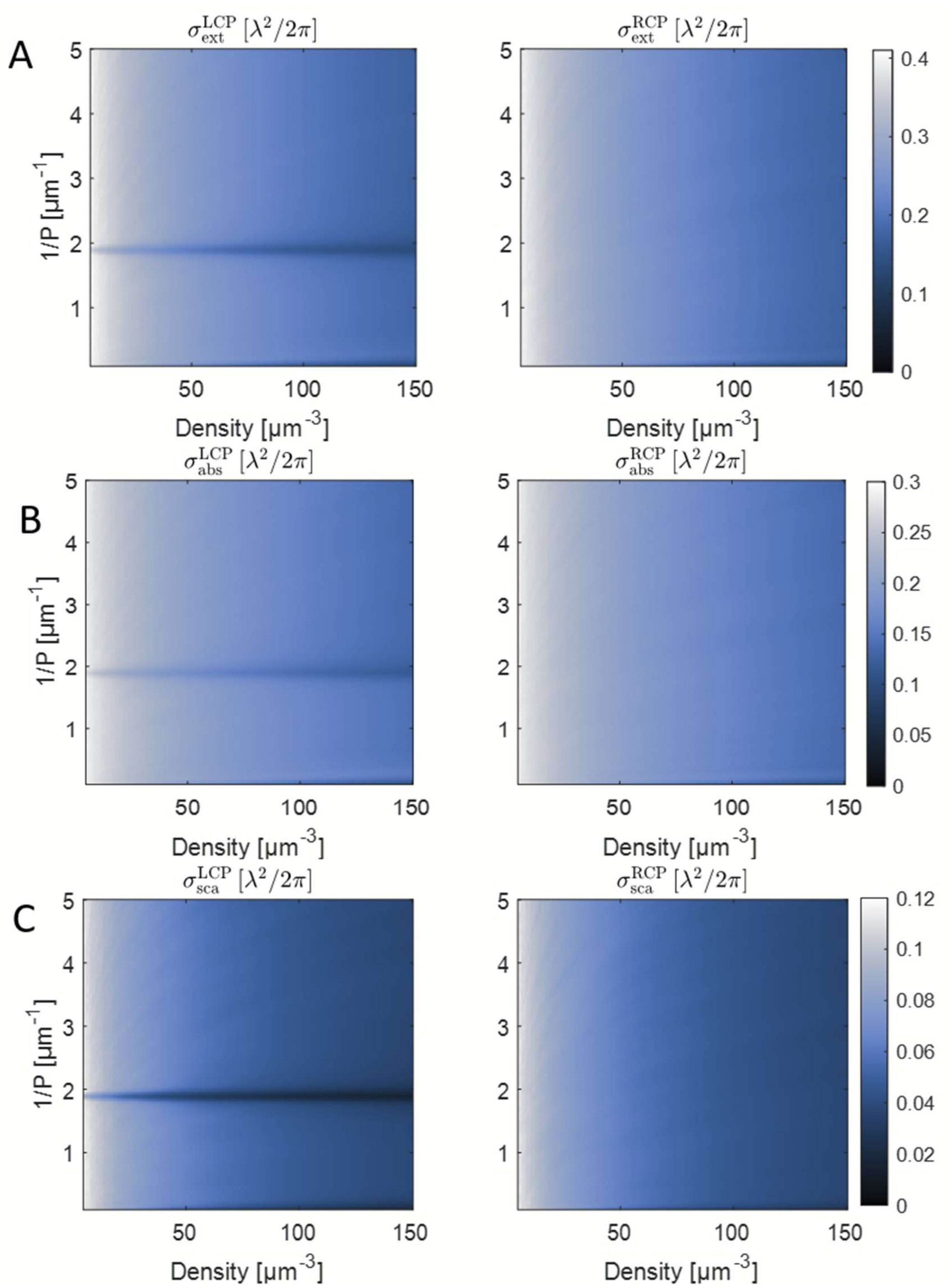

**Figure S25**. **Borrmann-type effect.** (**A**) Total extinction, (**B**) absorption, (**C**) and scattering cross sections of 1000 AuBPs in the *Helicoidal Model* under left- and right-circularly polarized illumination. The optical cross sections are largely insensitive to the helicoidal pitch over most of the parameter space. A pronounced deviation occurs when the pitch equals the wavelength inside the medium (n=1.5), where Bragg-like reflection leads to a selective suppression of absorption for one circular polarization. This behavior is consistent with a Borrmann-type effect and coincides with the handedness inversion boundary identified in Fig. 3E.

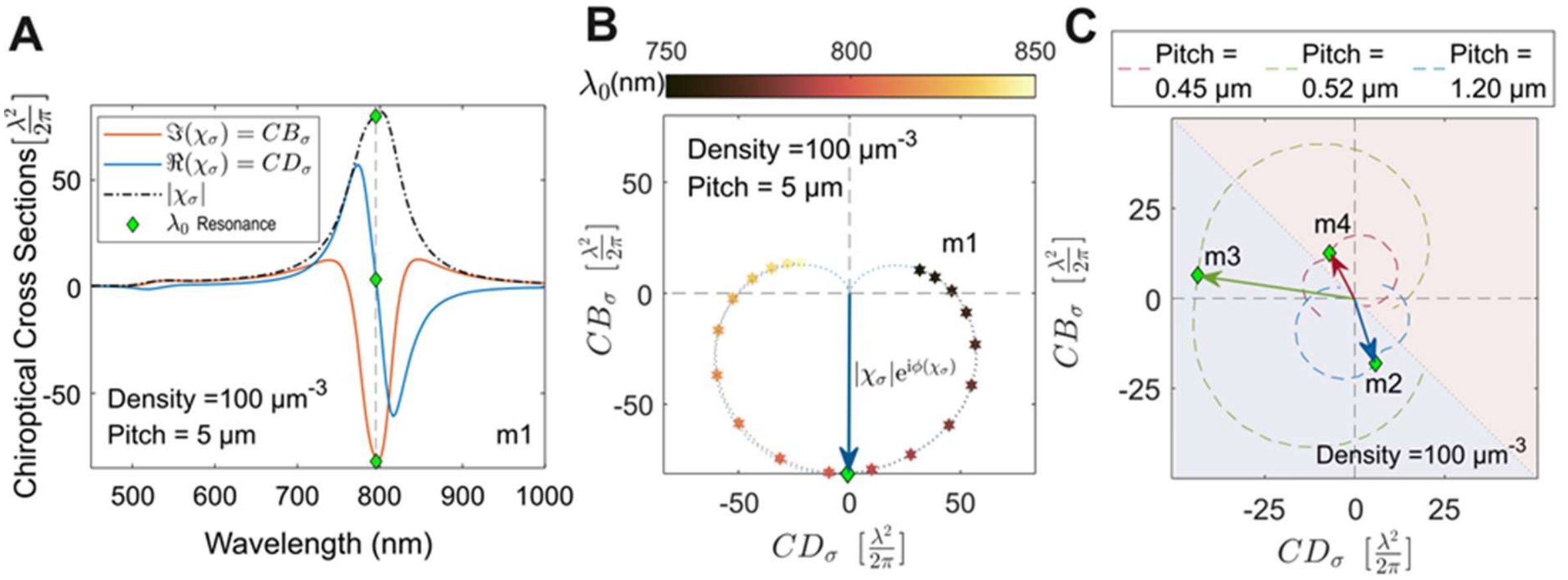

**Figure S26**. **Chiroptical index and chirality inversions at high densities**. **(A-B)** T-matrix simulation of the chiroptical index, $\chi_\sigma$, for sample m1 (1000 AuBPs, density=100 μm$^{-3}$, pitch = 5 μm, $n_{\text{Chitin}} = 1.5$) using the Helicoidal Model. (**A**) Spectra of the real part, Re($\chi_\sigma$), imaginary part, Im($\chi_\sigma$), and magnitude, $|\chi_\sigma|$. (**B**) The same data in the complex plane, where $\chi_\sigma$ traces a limaçon trajectory. The color bar indicates the clockwise progression of wavelength. In both panels, green diamonds indicate the plasmonic resonance wavelength. The arrow in (**B**) connects the origin to $\chi_\sigma$ at resonance, so its length and angle give the magnitude and phase of the index at the resonance. (**C**) Simulated chiroptical indices in the complex plane for samples m2, m3, and m4, which use the same parameters as m1 except for the pitch. Arrows point to the resonance wavelength. Tracking the resonance point from sample to sample shows the arrow rotating, which indicates the chirality inversion.